\documentclass[vecphys]{svmult}
\usepackage{makeidx}
\usepackage{graphicx}

\usepackage{multicol}
\usepackage[bottom]{footmisc}

\begin{document}

\title*{Particle Physics Approach to Dark
Matter}
\author{George Lazarides}
\institute{Physics Division, School of
Technology, Aristotle University of
Thessaloniki, Thessaloniki 54124, Greece
\texttt{lazaride@eng.auth.gr}}

\maketitle

\begin{abstract}
We review the main proposals of particle
physics for the composition of the cold dark
matter in the universe. Strong axion
contribution to cold dark matter is not
favored if the Peccei-Quinn field emerges
with non-zero value at the end of inflation
and the inflationary scale is superheavy since,
under these circumstances, it leads to
unacceptably large isocurvature perturbations.
The lightest neutralino is the most popular
candidate constituent of cold dark matter. Its
relic abundance in the constrained minimal
supersymmetric standard model can be reduced to
acceptable values by pole annihilation of
neutralinos or neutralino-stau
coannihilation. Axinos can also contribute to
cold dark matter provided that the reheat
temperature is adequately low. Gravitinos can
constitute the cold dark matter only in limited
regions of the parameter space. We present a
supersymmetric grand unified model leading to
violation of Yukawa unification and, thus,
allowing an acceptable $b$-quark mass within
the constrained minimal supersymmetric standard
model with $\mu>0$. The model possesses a wide
range of parameters consistent with the data on
the cold dark matter abundance as well as other
phenomenological constraints. Also, it leads to
a new version of shifted hybrid inflation.
\end{abstract}

\section{Introduction}
\label{sec:intro}

The recent measurements of the Wilkinson
microwave anisotropy probe (WMAP) satellite
\cite{wmap} on the cosmic microwave background
radiation (CMBR) have shown that the matter
abundance in the universe is
$\Omega_mh^2=0.135^{+0.008}_{-0.009}$, where
$\Omega_i=\rho_i/\rho_c$ with $\rho_i$ being
the energy density of the $i$-th species and
$\rho_c$ the critical energy density of the
universe and $h$ is the present value of the
Hubble parameter in units of
$100~{\rm km}~{\rm sec}^{-1}~{\rm Mpc}^{-1}$.
The baryon abundance is also found by these
measurements to be
$\Omega_bh^2=0.0224\pm 0.0009$. Consequently,
the cold dark matter (CDM) abundance in the
universe is $\Omega_{\rm CDM}h^2=
0.1126^{+0.00805}_{-0.00904}$. The $95\%$
confidence level (c.l.) range of
this quantity is then, roughly, $0.095
\stackrel{<}{_{\sim}}\Omega_{\rm CDM}h^2
\stackrel{<}{_{\sim}}0.13$. Taking
$h\simeq 0.72$, which is its best-fit value
from the Hubble space telescope (HST)
\cite{hst}, and assuming that the total energy
density of the universe is very close to its
critical energy density (i.e. $\Omega_{\rm tot}
\simeq 1$), as implied by WMAP, we conclude
that about $22\%$ of the energy density of the
present universe consists of CDM.

\par
The question then is, what the nature, origin,
and composition of this important component of
our universe is. Particle physics provides us
with a number of candidate particles out of
which CDM can be made. These particles
appear naturally in various particle physics
frameworks for reasons completely independent
from CDM considerations and are, certainly, not
invented for the sole purpose of explaining the
presence of CDM in the universe.

\par
The basic properties that such a candidate
particle must satisfy are the following: ($i$)
it must be stable or very long-lived, which can
be achieved by an appropriate symmetry, ($ii$)
it should be electrically and color neutral, as
implied by astrophysical constraints on exotic
relics (like anomalous nuclei), but can be
interacting weakly, and ($iii$) it has to be
non-relativistic, which is usually guaranteed
by assuming that it is adequately
massive, although even very light particles
such as axions can be non-relativistic for
different reasons. So, what we need as
constituent of CDM is a weakly interacting
massive particle. There are several
possibilities, but we will concentrate here on
the major particle physics candidates which are
the axion, the lightest neutralino, the axino,
and the gravitino (for other candidates, see
e.g. Ref.~\cite{other}). Note that the last
three particles exist only in supersymmetric
(SUSY) theories.

\par
In Sec.~\ref{sec:axion}, we examine the
possibility that the axions are constituents of
CDM. Sec.~\ref{sec:mssm} is devoted to
outlining the salient features of the minimal
supersymmetric standard model (MSSM), which
will be used as a basic frame for discussing
SUSY CDM. In Sec.~\ref{sec:relic}, we summarize
the calculation of the relic abundance of the
lightest neutralino, which is normally the
lightest supersymmetric particle (LSP), and
investigate the circumstances under which it
can account for the CDM in the universe. In
Secs.~\ref{sec:axino} and \ref{sec:gravitino},
we discuss, respectively, axinos and gravitinos
as constituents of CDM. In
Sec.~\ref{sec:quasi},
we present a SUSY grand unified theory (GUT)
model which solves the
bottom-quark mass problem by naturally and
modestly violating the exact unification of the
third generation Yukawa couplings. We study the
parameter space of the model which is allowed
by neutralino dark matter considerations as
well as some other phenomenological
constraints. Finally, in Sec.~\ref{sec:concl},
we summarize our conclusions.

\section{Axions}
\label{sec:axion}

\par
The most natural solution to the strong $CP$
problem (i.e. the apparent absence of $CP$
violation in strong interactions implied by
the experimental bound on the electric dipole
moment of the neutron) is the one provided by a
Peccei-Quinn (PQ) symmetry \cite{pq}. This is a
global ${\rm U}(1)$ symmetry,
${\rm U}(1)_{\rm PQ}$, which carries QCD
anomalies and is spontaneously broken at a
scale $f_a$, the so-called axion decay
constant or simply PQ scale. Astrophysical
\cite{astrop} and cosmological constraints
imply that $10^9~{\rm GeV}\stackrel{<}{_{\sim}}
f_a\stackrel{<}{_{\sim}}10^{12}~{\rm GeV}$. The
upper bound originates
\cite{overclosure1,overclosure2} from the
requirement that the relic energy density of
axions does not overclose the universe. It
should be noted, however, that this upper bound
can be considerably relaxed if the axions are
diluted \cite{overclosure2,dilution,relax} by
entropy generation after their production at
the QCD phase transition (for more recent
applications of this dilution mechanism, see
e.g. Ref.~\cite{dimop}).

\par
The axion is a pseudo Nambu-Goldstone boson
corresponding to the phase of the complex PQ
field, which breaks ${\rm U}(1)_{\rm PQ}$ by
its vacuum expectation value (VEV). After the
end of inflation \cite{inflation}, this phase
appears homogenized over the universe
(supposing that the PQ field is non-zero) with
a value $\theta$, which is known as the initial
misalignment angle. Naturalness suggests that
$\theta$ is of order unity. This angle remains
frozen until the QCD phase transition, where
the QCD instantons come into play. They break
explicitly the PQ symmetry to a discrete
subgroup \cite{sikivie} since this symmetry
carries QCD anomalies. So, a sinusoidal
potential for the phase of the PQ field is
generated and this phase starts oscillating
coherently about a minimum of the potential.
The resulting state resembles pressureless
matter consisting of static axions with mass
$m_a\sim\Lambda_{\rm QCD}^2/f_a$, where
$\Lambda_{\rm QCD}\sim 200~{\rm MeV}$ is the
QCD scale. For $f_a\sim 10^{12}~{\rm GeV}$, the
mass of the axion $m_a\sim 10^{-5}~{\rm eV}$.
Note that axions, although very light, are
good candidates for being constituents of the
CDM in the universe since they are produced at
rest. Also, they are very weakly interacting
since their interactions are suppressed by the
axion decay constant $f_a$.

\par
The relic abundance of axions can be calculated
by using the formulae of Ref.~\cite{turner},
where we take the QCD scale
$\Lambda_{\rm QCD}=200~{\rm MeV}$ and ignore
the uncertainties for simplicity. We find
\begin{equation}
\Omega_ah^2\approx\theta^2\left(\frac
{f_a}{10^{12}~{\rm GeV}}\right)^{1.175}
\label{axionabundance}
\end{equation}
(note that a primordial magnetic helicity, may
\cite{magnetic} influence this abundance). So,
for natural values of $\theta\sim 0.1$ and
$f_a\sim 10^{12}~{\rm GeV}$, axions can
contribute significantly to CDM, which can
even consist solely of axions.

\par
The main disadvantage of axionic dark matter is
that it leads to isocurvature perturbations if
the PQ field emerges with non-zero
(homogeneous) value at the end of inflation.
Indeed, during inflation, the angle $\theta$
acquires a superhorizon spectrum
of perturbations as all the almost massless
degrees of freedom. At the QCD phase
transition, these perturbations turn into
isocurvature perturbations in the axion energy
density, which means that the partial curvature
perturbation in axions is different than the
one in photons. The recent results of WMAP
\cite{wmap} put stringent bounds
\cite{wmapinf,iso,trotta} on the possible
isocurvature perturbation. So, a large axion
contribution to CDM is disfavored in models
where the inflationary scale is superheavy
(i.e. of the order of the SUSY GUT scale) and
the PQ field is non-zero at the end of
inflation.

\par
We now wish to turn to the discussion of the
main SUSY candidates for dark matter: the
lightest neutralino $\tilde\chi$, the axino
$\tilde{a}$ and the gravitino $\tilde{G}$. We
will consider them mainly within the simplest
SUSY framework, which is the MSSM. It is,
thus, important to first outline the basics of
MSSM.

\section{Salient Features of MSSM}
\label{sec:mssm}

\par
We consider the MSSM embedded in some general
SUSY GUT model. We further assume that the GUT
gauge group breaking down to the standard model
(SM) gauge group $G_{\rm SM}$ occurs in one
step at a scale $M_{\rm GUT}\sim 10^{16}~
{\rm GeV}$, where the gauge coupling constants
of strong, weak, and electromagnetic
interactions unify. Ignoring the Yukawa
couplings of the first and second generation,
the effective superpotential below
$M_{\rm GUT}$ is
\begin{equation}
W=\epsilon_{ij}(-h_t H_2^i q_3^j t^c+
h_b H_1^i q_3^j b^c+h_\tau H_1^i l_3^j\tau^c-
\mu H_1^i H_2^j),
\label{super}
\end{equation}
where $q_3=(t,b)$ and $l_3=(\nu_{\tau},\tau)$
are the quark and lepton ${\rm SU}(2)_{\rm L}$
doublet left handed superfields of the third
generation and $t^c$, $b^c$, and $\tau^c$ the
corresponding ${\rm SU}(2)_{\rm L}$ singlets.
Also, $H_1$, $H_2$ are the electroweak Higgs
superfields and $\epsilon$ the $2\times 2$
antisymmetric matrix with $\epsilon_{12}=+1$.
The gravity-mediated soft SUSY-breaking terms
in the scalar potential are given by
$$
V_{\rm soft}=\sum_{a,b}m_{ab}^{2}\phi^{*}_a
\phi_b+
$$
\begin {equation}
\left[\epsilon_{ij}(-A_th_tH_2^i\tilde q_3^j
\tilde t^c+A_b h_b H_1^i\tilde q_3^j\tilde b^c
+A_\tau h_\tau H_1^i \tilde l_3^j\tilde\tau^c
-B\mu H_1^iH_2^j)+ {\rm h.c.}\right],
\label{vsoft}
\end{equation}
where the sum is taken over all the complex
scalar fields $\phi_a$ and tildes denote
superpartners. The soft gaugino mass terms in
the Lagrangian are
\begin{equation}
\mathcal{L}_{\rm gaugino}=\frac{1}{2}\left(
M_1\tilde B\tilde B+M_2\sum_{r=1}^{3}\tilde W_r
\tilde W_r+M_3\sum_{a=1}^{8}\tilde g_a\tilde
g_a+{\rm h.c.}\right),
\label{gaugino}
\end{equation}
where $\tilde B$, $\tilde W_r$ and $\tilde g_a$
are the bino, winos and gluinos respectively.

\par
The Lagrangian of MSSM is invariant under a
discrete $Z_2$ matter parity symmetry under
which all ``matter'' (i.e. quark and lepton)
superfields change sign. Combining this
symmetry with the $Z_2$ fermion number symmetry
under which all fermions change sign, we obtain
the discrete $Z_2$ R-parity symmetry under
which all SM particles are
even, while all sparticles are odd. By virtue
of R-parity conservation, the LSP is stable
and, thus, can contribute to the CDM in the
universe. It is important to note that matter
parity is vital for MSSM to avoid baryon-
and lepton-number-violating
renormalizable couplings in the superpotential,
which would lead to highly undesirable
phenomena such as very fast proton decay. So,
the possibility of having the LSP as CDM
candidate is not put in by hand, but arises
naturally from the very structure of MSSM.

\par
The SUSY-breaking parameters
$m_{ab},A_t,A_b,A_\tau,B$, and $M_i$
($i=1,2,3$) are all of the order
of the soft SUSY-breaking scale $M_{\rm SUSY}
\sim 1~{\rm TeV}$, but are otherwise unrelated
in the general case. However, if we assume that
soft SUSY breaking is mediated by minimal
supergravity (mSUGRA), i.e. supergravity with
minimal K\"{a}hler potential, we
obtain soft terms which are universal
``asymptotically'' (i.e. at $M_{\rm GUT}$). In
particular, we obtain a common scalar mass
$m_0$, a common trilinear scalar coupling
$A_0$, and a common gaugino mass $M_{1/2}$. The
MSSM supplemented by universal boundary
conditions is known as constrained MSSM
(CMSSM) \cite{Cmssm}. It is true that mSUGRA
implies two more asymptotic relations:
$B_0=A_0-m_0$ and $m_0=m_{3/2}$, where
$B_0=B(M_{\rm GUT})$ and $m_{3/2}$ is the
(asymptotic) gravitino mass. These extra
conditions are usually not included in the
CMSSM. Imposing them, we get the so-called
very CMSSM \cite{vCmssm}, which is a very
restrictive version of MSSM and will not be
considered in these lectures.

\par
The CMSSM can be further restricted by imposing
asymptotic Yukawa unification (YU)
\cite{als}, i.e. the equality of all three
Yukawa coupling constants of the third family
at $M_{\rm GUT}$:
\begin{equation}
h_t(M_{\rm GUT})=h_b(M_{\rm GUT})=
h_{\tau}(M_{\rm GUT})\equiv h_0.
\label{yukawa}
\end{equation}
Exact YU, which makes the CMSSM considerably
more predictive, can be obtained in GUTs based
on a gauge group such as ${\rm SO}(10)$ or
${\rm E}_{6}$ under which all the particles of
one family belong to a single representation
with the additional requirement that the masses
of the third family fermions arise primarily
from their unique Yukawa coupling to a single
superfield representation which predominantly
contains the electroweak Higgs superfields. It
should be noted that exact YU in the CMSSM
leads
to unacceptable values for the bottom-quark
mass $m_b$ and, thus, must be corrected in
order to become consistent with observations.
We will ignore this problem for the moment, but
we will return to it in Sec.~\ref{sec:quasi}.

\par
Now, we assume that our effective theory below
$M_{\rm GUT}$ is the CMSSM with YU. This theory
depends on the following parameters
($\mu_0=\mu(M_{\rm GUT})$):
\[
m_0,\ M_{1/2},\ A_0,\ \mu_0,\ B_0,\
\alpha_{\rm GUT},\ M_{\rm GUT},\ h_{0},\
\tan\beta,\]
where
$\alpha_{\rm GUT}\equiv g_{\rm GUT}^{2}/4\pi$
with $g_{\rm GUT}$ being the GUT gauge coupling
constant and $\tan\beta\equiv\langle H_2\rangle
/\langle H_1\rangle$ is the ratio of the two
electroweak VEVs.
The parameters $\alpha_{\rm GUT}$ and
$M_{\rm GUT}$ are evaluated consistently with
the experimental values of the electromagnetic
and strong fine-structure constants
$\alpha_{\rm em}$ and $\alpha_s$, and the
sine-squared of the Weinberg angle
$\sin^2\theta_W$ at $M_Z$. To this end, we
integrate \cite{cdm} numerically the
renormalization group equations (RGEs) for the
MSSM at two loops in the gauge and Yukawa
coupling constants from $M_{\rm GUT}$ down to a
common but variable \cite{threshold} SUSY
threshold $M_{\rm SUSY}\equiv
\sqrt{m_{\tilde{t}_1} m_{\tilde{t}_2}}$
($\tilde{t}_{1,2}$ are the stop-quark mass
eigenstates). From $M_{\rm SUSY}$ to $M_Z$, the
RGEs of the non-SUSY SM are used. The
set of RGEs needed for our computation can be
found in many references (see e.g.
Ref.~\cite{rge}). We take
$\alpha_s(M_Z)=0.12\pm 0.001$ which, as it
turns out, leads to gauge coupling unification
at $M_{\rm GUT}$ with an accuracy better than
0.1$\%$. So, we can assume exact unification
once the appropriate SUSY particle thresholds
are taken into account.

\par
The unified third generation Yukawa coupling
constant $h_0$ at $M_{\rm GUT}$ and the value
of $\tan\beta$ at $M_{\rm SUSY}$ are estimated
using the experimental inputs for the top-quark
mass $m_t(m_t)=166~{\rm GeV}$ and the
$\tau$-lepton mass
$m_\tau(M_Z)=1.746~{\rm{ GeV}}$. Our
integration procedure of the RGEs relies
\cite{cdm} on iterative runs of these equations
from $M_{\rm GUT}$ to low energies and back for
every set of values of the input parameters
until agreement with the experimental data is
achieved. The SUSY corrections to $m_\tau$ are
taken from Ref.~\cite{pierce} and incorporated
at $M_{\rm SUSY}$.

\par
Assuming radiative electroweak symmetry
breaking, we can express the values of the
parameters $\mu$ (up to its sign) and $B$ (or,
equivalently, the mass $m_A$ of the $CP$-odd
neutral Higgs boson $A$) at $M_{\rm SUSY}$ in
terms of the other input parameters by
minimizing the tree-level renormalization
group (RG) improved potential \cite{grz} at
$M_{\rm SUSY}$. The resulting conditions are
\begin{equation}
\mu^2=\frac{m^2_{H_1}-m^2_{H_2}\tan^2{\beta}}
{\tan^2{\beta}-1}-\frac{1}{2} M^2_Z , \quad
\sin 2\beta=\frac{2 B\mu}{m_{H_1}^2+m_{H_2}^2+
2 \mu^2}\equiv \frac{2B\mu}{m_A^2},
\label{mu}
\end{equation}
where $m_{H_1}$, $m_{H_2}$ are the soft
SUSY-breaking scalar Higgs masses. We can
improve the accuracy of these conditions by
including the full one-loop radiative
corrections to the potential from
Ref.~\cite{pierce} at $M_{\rm SUSY}$. We
find that the corrections to $\mu$ and $m_A$
from the full one-loop effective potential
are minimized \cite{threshold,bcdpt} by our
choice of $M_{\rm SUSY}$. So, a much better
accuracy is achieved by using this variable
SUSY threshold rather than a fixed one.
Furthermore,
we include in our calculation the two-loop
radiative corrections to the masses $m_h$ and
$m_H$ of the $CP$-even neutral Higgs bosons $h$
and $H$. These corrections are particularly
important for the mass of the lightest
$CP$-even neutral Higgs boson $h$. Finally, the
SUSY corrections to $m_b$ are also included at
$M_{\rm SUSY}$ using the relevant formulae of
Ref.~\cite{pierce}. As already mentioned, the
predicted value of the bottom-quark mass is not
compatible with experiment. However, we will
ignore this problem for the moment. The sign of
$\mu$ is taken to be positive, since the
$\mu<0$ case is excluded because it leads
\cite{borzumati,bbct} to a neutralino relic
abundance which is well above unity, thereby
overclosing the universe, for all $m_A$'s
permitted by $b\rightarrow s\gamma$. We are
left with $m_0,\ M_{1/2}$ and $A_0$ as
free input parameters.

\par
The LSP is the lightest neutralino
$\tilde{\chi}$. The mass matrix for the four
neutralinos is
\begin{eqnarray}
\left(\matrix{
M_1 & 0 & -M_Z s_W\cos\beta & M_Z s_W\sin\beta
\vspace{0.3cm} \cr
0 & M_2 & M_Z c_W\cos\beta & -M_Z c_W\sin\beta
\vspace{0.3cm} \cr
-M_Z s_W\cos\beta & M_Z c_W\cos\beta & 0 & -\mu
\vspace{0.3cm} \cr
M_Z s_W\sin\beta & -M_Z c_W\sin\beta & -\mu & 0
\cr}\right)
\label{neutralino}
\end{eqnarray}
in the $(-i\tilde B, -i\tilde W_3, \tilde H_1,
\tilde H_2)$ basis. Here, $s_W=\sin \theta_W$,
$c_W=\cos \theta_W $, and $M_1$, $M_2$ are the
mass parameters of $\tilde B$, $\tilde W_3$ in
Eq.~(\ref{gaugino}). In CMSSM, the lightest
neutralino turns out to be an almost pure bino
$\tilde B$.

\par
The LSPs are stable due to the presence of the
unbroken R-parity, but can annihilate in pairs
since this symmetry is a discrete $Z_2$
symmetry. This reduces their relic abundance in
the universe. If there exist
sparticles with masses close to the mass of the
LSP, their coannihilation \cite{coan} with the
LSP leads to a further reduction of the LSP
relic abundance. It should be noted that the
number density of these sparticles is not
Boltzmann suppressed relative to the LSP number
density. They eventually decay yielding an
equal number of LSPs and, thus, contributing to
the relic abundance of the LSPs. Of
particular importance is the next-to-LSP
(NLSP), which, in CMSSM, turns out to be the
lightest stau mass eigenstate $\tilde\tau_2$.
Its mass is obtained by diagonalizing the stau
mass-squared matrix
\begin{eqnarray}
\left(\matrix{
m_{\tau}^2+m_{\tilde\tau_{\rm L}}^2+M_Z^2
(-\frac{1}{2}+s_W^2)\cos 2\beta
& m_{\tau}(A_{\tau}-\mu\tan\beta)
\vspace{0.3cm} \cr
m_{\tau}(A_{\tau}-\mu\tan\beta)
& m_{\tau}^2+m_{\tilde\tau_{\rm R}}^2-M_Z^2
s_W^2\cos 2\beta\cr}
\right)
\label{stau}
\end{eqnarray}
in the gauge basis
($\tilde\tau_{\rm L},\ \tilde\tau_{\rm R}$).
Here, $m_{\tilde\tau_{\rm L[R]}}$ is the soft
SUSY-breaking mass of the left [right] handed
stau $\tilde\tau_{\rm L[R]}$ and $m_{\tau}$ the
tau-lepton mass. The stau mass eigenstates are
\begin{eqnarray}
\left( \matrix{
\tilde\tau_1
\vspace{0.3cm} \cr
\tilde\tau_2
\cr}
\right)
& = &
\left( \matrix{
\cos\theta_{\tilde{\tau}} &
\sin\theta_{\tilde{\tau}}
\vspace{0.3cm}\cr
-\sin\theta_{\tilde{\tau}} &
\cos\theta_{\tilde{\tau}}
\cr}
\right)
\left( \matrix{
\tilde\tau_{\rm L}
\vspace{0.3cm}\cr
\tilde\tau_{\rm R}
\cr}
\right),
\label{eigen}
\end{eqnarray}
where $\theta_{\tilde{\tau}}$ is the
$\tilde\tau_{\rm L}-\tilde\tau_{\rm R}$
mixing angle.

\par
The large values of $b$ and $\tau$ Yukawa
coupling constants implied by YU cause soft
SUSY-breaking masses of the third generation
squarks and sleptons to run (at low energies)
to lower physical values than the
corresponding masses of the first and second
generation. Furthermore, the large values of
$\tan\beta$ implied by YU lead to large
off-diagonal mixings in the sbottom and stau
mass-squared matrices. These effects reduce
further the physical mass of the lightest
stau, which is the NLSP. Another effect of
the large values of the $b$ and $\tau$ Yukawa
coupling constants is the reduction of the mass
$m_A$ of the $CP$-odd neutral Higgs boson $A$
and, consequently, the other Higgs boson masses
to smaller values.

\section{Neutralino Relic Abundance}
\label{sec:relic}

We now turn to the calculation of the
cosmological relic abundance of the lightest
neutralino $\tilde\chi$ (almost pure
$\tilde B$) in the CMSSM with YU according to
the standard cosmological scenario (for
non-standard scenaria, see e.g.
Ref.~\cite{pallis}). In general, all sparticles
contribute to $\Omega_{\tilde\chi}h^2$, since
they eventually turn into LSPs, and all the
(co)annihilation processes must be considered.
The most important contributions, however, come
from the LSP and the NLSP. So, in the case of
the CMSSM, we should concentrate on
$\tilde\chi$ (LSP) and $\tilde{\tau}_2$ (NLSP)
and consider the coannihilation of $\tilde\chi$
with $\tilde\tau_2$ and $\tilde\tau_2^\ast$.
The important role of the coannihilation of the
LSP with sparticles carrying masses close to
its mass in the calculation of the LSP relic
abundance has been pointed out by many authors
(see e.g.
Refs.~\cite{cdm,coan,drees,ellis1,ellis2}).
Here, we will use the method of
Ref.~\cite{coan}, which was also used in
Ref.~\cite{cdm}. Note that our analysis can be
readily applied to any MSSM scheme where the
LSP and NLSP are the bino and stau
respectively. In particular, it applies to the
CMSSM without YU, where we have $\tan\beta$
as an extra free input parameter.

\par
The relevant quantity, in our case, is the
total number density
\begin{equation}
n= n_{\tilde\chi} + n_{\tilde\tau_2}
+ n_{\tilde\tau_2^\ast},
\label{n}
\end{equation}
since the $\tilde\tau_2$'s and
$\tilde\tau_2^\ast$'s decay into $\tilde\chi$'s
after freeze-out. At cosmic temperatures
relevant for freeze-out, the scattering rates
of these (non-relativistic) sparticles off
particles in the thermal bath are much faster
than their annihilation rates since the
(relativistic) particles in the bath are
considerably more abundant. Consequently, the
number densities $n_i$ ($i=\tilde\chi$,
$\tilde\tau_2$, $\tilde\tau_2^\ast$) are
proportional to their equilibrium values
$n_i^{\rm eq}$ to a good approximation, i.e.
$n_i/n\approx n_i^{\rm eq}/n^{\rm eq}\equiv
r_i$. The Boltzmann equation (see e.g.
Ref.~\cite{kt}) is then written as
\begin{equation}
\frac{dn}{dt}=-3Hn-\langle \sigma_{\rm eff}v
\rangle(n^2-(n^{\rm eq})^2),
\label{boltzmann}
\end{equation}
where $H$ is the Hubble parameter, $v$ is the
``relative velocity'' of the annihilating
particles, $\langle\cdot\cdot\cdot\rangle$
denotes thermal averaging and
$\sigma_{\rm eff}$ is the effective
cross section defined by
\begin{equation}
\sigma_{\rm eff}=\sum_{i,j}\sigma_{ij}r_ir_j
\label{sigmaeff1}
\end{equation}
with $\sigma_{ij}$ being the total
cross section for particle $i$ to annihilate
with particle $j$ averaged over initial spin
states. In our case, $\sigma_{\rm eff}$ takes
the following form
\begin{equation}
\sigma_{\rm eff}=
\sigma_{\tilde\chi\tilde\chi}
r_{\tilde\chi}r_{\tilde\chi}+
4\sigma_{\tilde\chi\tilde\tau_2}
r_{\tilde\chi}r_{\tilde\tau_2}+
2(\sigma_{\tilde\tau_2\tilde\tau_2}+
\sigma_{\tilde\tau_2\tilde\tau_2^\ast})
r_{\tilde\tau_2}r_{\tilde\tau_2}.
\label{sigmaeff2}
\end{equation}
For $r_i$, we use the non-relativistic
approximation
\begin{equation}
r_i(x)=\frac{g_i(1+\Delta_i)^{\frac{3}{2}}
e^{-\Delta_i x}}{g_{\rm eff}},
\label{ri}
\end{equation}
\begin{equation}
g_{\rm eff}(x)={\sum_{i}g_i
(1+\Delta_i)^{\frac{3}{2}}e^{-\Delta_i x}},
\quad\Delta_i=\frac{m_i-m_{\tilde\chi}}
{m_{\tilde\chi}}.
\label{geff}
\end{equation}
Here $g_i=2$, 1, 1 ($i=\tilde\chi$,
$\tilde\tau_2$, $\tilde\tau_2^\ast$) is the
number of degrees of freedom of the $i$-th
particle with mass $m_i$ and
$x=m_{\tilde\chi}/T$ with $T$ being the photon
temperature.

\par
Using Boltzmann equation (which is depicted in
Eq.~(\ref{boltzmann})), we can calculate the
relic abundance of the LSP at the present
cosmic time. It has been found \cite{coan,kt}
to be given by
\begin{equation}
\Omega_{\tilde\chi}h^2\approx\frac{1.07
\times 10^9~{\rm GeV}^{-1}}
{g_*^{\frac{1}{2}}M_{\rm P}\,x_F^{-1}
\hat\sigma_{\rm eff}}
\label{omega}
\end{equation}
with
\begin{equation}
\hat\sigma_{\rm eff}\equiv x_F
\int_{x_F}^{\infty}
\langle\sigma_{\rm eff}v\rangle x^{-2}dx.
\label{sigmaeff3}
\end{equation}
Here $M_{\rm P}\simeq 1.22 \times 10^{19}~
{\rm GeV}$ is the Planck scale, $g_*\simeq 81$
is the effective number of massless degrees of
freedom at freeze-out \cite{kt} and
$x_F=m_{\tilde\chi}/T_{F}$ with $T_F$ being the
freeze-out photon temperature calculated by
solving iteratively the equation \cite{kt,gkt}
\begin{equation}
x_F=\ln\frac{0.038\,g_{\rm eff}(x_F)\,M_{\rm P}
\,(c+2)\,c\,m_{\tilde\chi}\,\langle
\sigma_{\rm eff}v\rangle (x_F)}
{g_*^{\frac{1}{2}}\,x_F^{\frac{1}{2}}}.
\label{xf}
\end{equation}
The constant $c$ is chosen to be equal to $1/2$
\cite{gkt}. The freeze-out temperatures which
we obtain here are of the order of
$m_{\tilde\chi}/25$ and, thus, our
non-relativistic approximation (see
Eq.~(\ref{ri})) is {\it a posteriori}
justified.

\par
Away from s-channel poles and final-state
thresholds, the quantities $\sigma_{ij}v$ are
well approximated by applying the
non-relativistic Taylor expansion up to second
order in the relative velocity $v$:
\begin{equation}
\sigma_{ij}v=a_{ij}+b_{ij} v^2.
\label{taylorv}
\end{equation}
Actually, this corresponds \cite{drees} to an
expansion in s and p waves. The thermally
averaged cross sections are then easily
calculated
\begin{equation}
\langle\sigma_{ij}v\rangle (x)=
\frac{x^{\frac{3}{2}}}{2\sqrt{\pi}}
\int_{0}^{\infty}dvv^2(\sigma_{ij}v)
e^{-\frac{xv^2}{4}}=a_{ij}+6 \frac{b_{ij}}{x}.
\label{average}
\end{equation}
Here, we approximated the masses of the
incoming particles by the neutralino mass, i.e.
$m_i=m_j=m_{\tilde{\chi}}$. The reduced mass
of the incoming particles is then equal to
$m_{\tilde{\chi}}/2$. We also thermally
averaged over the relative velocity rather than
the separate velocities of the incoming
particles, which would be more accurate. Using
Eqs.~(\ref{sigmaeff1}), (\ref{sigmaeff2}),
(\ref{sigmaeff3}), and (\ref{average}), one
obtains
\begin{equation}
\hat\sigma_{\rm eff}=\sum_{(ij)}(\alpha_{(ij)}
a_{ij}+\beta_{(ij)}b_{ij})\equiv
\sum_{(ij)}\hat\sigma_{(ij)},
\label{sigmaeff4}
\end{equation}
where we sum over $(ij)=(\tilde\chi\tilde\chi)$,
$(\tilde\chi\tilde\tau_2)$, and $(\tilde\tau_2
\tilde\tau_2^{(\ast)})$ with
$a_{\tilde\tau_2\tilde\tau_2^{(\ast)}}=
a_{\tilde\tau_2\tilde\tau_2}+
a_{\tilde\tau_2\tilde\tau_2^{\ast}}$,
$b_{\tilde\tau_2\tilde\tau_2^{(\ast)}}=
b_{\tilde\tau_2\tilde\tau_2}+
b_{\tilde\tau_2\tilde\tau_2^{\ast}}$, and
$\alpha_{(ij)}$, $\beta_{(ij)}$ given by
\begin{equation}
\alpha_{(ij)}=c_{(ij)}x_F\int_{x_F}^\infty
\frac{dx}{x^2}r_i(x)r_j(x),\quad
\beta_{(ij)}=6c_{(ij)}x_F\int_{x_F}^\infty
\frac{dx}{x^3}r_i(x)r_j(x).
\label{alphabeta}
\end{equation}
Here $c_{(ij)}=1$, 4, 2 for $(ij)=
(\tilde\chi\tilde\chi)$,
$(\tilde\chi\tilde\tau_2)$, and
$(\tilde\tau_2\tilde\tau_2^{(\ast)})$
respectively.

\begin{table}
\centering
\caption{Feynman Diagrams}
{\footnotesize
\begin{tabular}{|c|c|c|}
\hline {} & {} & {} \\ [-2.2ex]
\multicolumn{1}{|c|}{Initial State} &
\multicolumn{1}{|c|}{Final States} &
\multicolumn{1}{|c|}{Diagrams}
\\ [0.4ex]
\hline {} & {} & {} \\ [-2.2ex]
$\tilde\chi\tilde\chi$ & $f\bar{f}$ &
$s(h,H,A,Z),~t(\tilde{f}),
~u(\tilde{f})$
\\ [0.4ex]
& $~hh,~hH,~HH,~HA,~AA,~ZA,~ZZ~$ &
$s(h,H),~t(\tilde\chi),~u(\tilde\chi)$
\\ [0.4ex]
& $hA,~hZ,~HZ$ & $s(A,Z),~t(\tilde\chi),
~u(\tilde\chi)$
\\ [0.4ex]
& $H^+H^-,~W^+W^-$ & $s(h,H,Z),
~t(\tilde\chi^\pm),~u(\tilde\chi^\pm)$
\\ [0.4ex]
& $W^\pm H^\mp $ & $s(h,H,A),
~t(\tilde\chi^\pm),~u(\tilde\chi^\pm)$
\\ [0.4ex]
\hline {} & {} & {} \\ [-2.2ex]
$\tilde\chi\tilde\tau_2$ &
$\tau h,~\tau H,~\tau Z$ &
$s(\tau),~t(\tilde\tau_{1,2})$
\\ [0.4ex]
& $\tau A$ & $s(\tau),~t(\tilde\tau_1)$
\\ [0.4ex]
& $\tau\gamma$ & $s(\tau),~t(\tilde\tau_2)$
\\ [0.4ex]
\hline {} & {} & {} \\ [-2.2ex]
$\tilde\tau_2\tilde\tau_2$ &
$\tau\tau$ & $t(\tilde\chi),~u(\tilde\chi)$
\\ [0.4ex]
\hline {} & {} & {} \\ [-2.2ex]
$\tilde\tau_2\tilde\tau_2^\ast$ &
$hh,~hH,~HH,~ZZ$ & $~s(h,H),
~t(\tilde\tau_{1,2}),~u(\tilde\tau_{1,2}),~c~$
\\ [0.4ex]
& $AA$ & $s(h,H),~t(\tilde\tau_1),
~u(\tilde\tau_1),~c$
\\ [0.4ex]
& $H^+ H^-,~W^+ W^-$ & $s(h,H,\gamma,Z),
~t(\tilde\nu_{\tau}),~c$
\\ [0.4ex]
& $\gamma\gamma,~\gamma Z$ & $t(\tilde\tau_2),
~u(\tilde\tau_2),~c$
\\ [0.4ex]
& $t\bar t,~b\bar b$ & $s(h,H,\gamma,Z)$
\\ [0.4ex]
& $\tau\bar\tau$ & $s(h,H,\gamma,Z),
~t(\tilde\chi)$
\\ [0.4ex]
& $u\bar u,~d\bar d,~e \bar e$ & $s(\gamma,Z)$
\\ [0.4ex]
\hline
\end{tabular}}
\label{tab:graphs}
\end{table}

\par
It should be emphasized that, near s-channel
poles or final-state thresholds, the Taylor
expansion in Eq.~(\ref{taylorv}) fails
\cite{coan,pole} badly and, thus, the thermal
average in Eq.~(\ref{average}) has to be
calculated accurately by numerical methods.
Also, for better accuracy, we should use fully
relativistic formulae instead of the
non-relativistic expressions in
Eqs.~(\ref{sigmaeff2}), (\ref{ri}), and
(\ref{average}). Finally, in
Eq.~(\ref{average}), we must take the
thermal average over the two initial particle
velocities $v_i$ and $v_j$ separately and not
just over their relative velocity $v$. The
masses of the incoming particles should also
be taken different $m_i\neq m_j$. After all
these improvements, Eq.~(\ref{average}) takes
\cite{gondolo} the form
\begin{equation}
\langle\sigma_{ij}v\rangle=\frac{1}
{2m_i^2m_j^2TK_2\left(\frac{m_i}{T}\right)
K_2\left(\frac{m_j}{T}\right)}
\int_{(m_i+m_j)^2}^\infty ds\,
K_1\left(\frac{\sqrt{s}}{T}\right)p_{ij}^2(s)
\sqrt{s}\,\sigma_{ij}(s),
\label{corraverage}
\end{equation}
where $K_n$ are Bessel functions, $s$ the
usual Mandelstam variable,
\begin{equation}
p_{ij}^2(s)=\frac{s}{4}-\frac{m_i^2+m_j^2}{2}+
\frac{(m_i^2-m_j^2)^2}{4s},
\label{pi2}
\end{equation}
and
\begin{equation}
\sigma_{ij}(s)=\frac{1}{4\sqrt{s}p_{ij}(s)}
\int\frac{d^3p^\prime}{(2\pi)^3E^\prime}
\frac{d^3p^{\prime\prime}}
{(2\pi)^3E^{\prime\prime}} (2\pi)^4
\delta^4(p_i+p_j-p^{\prime}-p^{\prime\prime})
|\mathcal{T}_{ij}|^2
\label{corrsigma}
\end{equation}
with $p^{\prime}$, $p^{\prime\prime}$,
$E^{\prime}$, $E^{\prime\prime}$ being the
3-momenta and energies of the outgoing
particles and $|\mathcal{T}_{ij}|^2$ the
squared transition matrix element summed
over final-state spins and averaged over
initial-state spins. Summation over all
final states is implied.

\par
The relevant final states and Feynman diagrams
for $\tilde{\chi}-\tilde{\tau}_2$
(co)annihilation are listed in
Table~\ref{tab:graphs}. The exchanged particles
are indicated for each pair of initial and
final states. The symbols $s(x,y,...)$,
$t(x,y,...)$, and $u(x,y,...)$ denote
tree-level graphs in which the particles
$x,y,...$ are exchanged in the s-, t-, and
u-channel respectively. The symbol $c$ stands
for ``contact'' diagrams with all four external
legs meeting at a vertex. The charged Higgs
bosons are denoted as $H^\pm$, while $f$
stands for all the matter fermions (quarks and
leptons) and $e$, $u$, and $d$ represent the
first and second generation charged leptons,
up-, and down-type quarks respectively. The
bars denote the anti-fermions,
$\tilde{\chi}^\pm$ are the charginos, and
$\tilde{\nu}_\tau$ is the superpartner of the
$\tau$-neutrino. We have included all possible
$\tilde{\chi}-\tilde{\chi}$ annihilation
processes (see e.g. Ref.~\cite{nihei1}), but
only the most important
$\tilde{\chi}-\tilde{\tau}_2$,
$\tilde{\tau}_2-\tilde{\tau}_2$, and
$\tilde{\tau}_2-\tilde{\tau}_2^\ast$
coannihilation processes from
Refs.~\cite{cdm,qcdm} (for a complete list see
e.g. Ref.~\cite{nihei2}), which are though
adequate for giving accurate results for
all values of $\tan\beta$, including the large
ones. Some of the diagrams listed here have not
been considered in previous works
\cite{ellis1,ellis2} with small $\tan\beta$.

\par
The $\tilde{\chi}-\tilde{\chi}$ annihilation
via an $A$- or $H$-pole exchange in the
s-channel can be \cite{bargerkao} very
important especially in the CMSSM with large
$\tan\beta$. As $\tan\beta$ increases, the
Higgs boson masses $m_A$ and $m_H$ decrease due
to the fact that $h_b$ increases and, thus, its
influence on the RG running of these masses is
enhanced. As a consequence, $m_A$ and $m_H$
approach $2m_{\tilde{\chi}}$ and the
neutralino pair annihilation via an $A$- or
$H$-pole exchange in the s-channel is
resonantly enhanced. The contribution from the
$H$ pole is p-wave suppressed as one can show
\cite{drees} using $CP$ invariance (recall that
the p wave is suppressed by $x_F\sim 25$).
Therefore, the dominant contribution originates
from the $A$ pole with the dominant decay mode
being the one to $b\bar{b}$ since, for large
$\tan\beta$, the $Ab\bar{b}$ coupling is
enhanced. We find \cite{funnel} that there
exists a region in the parameter space of the
CMSSM corresponding to large values of
$\tan\beta$ where the
$\tilde{\chi}-\tilde{\chi}$ annihilation via an
$A$ pole reduces drastically the relic
neutralino abundance and, thus, makes it
possible to satisfy the WMAP constraint on CDM
(note that, generically,
$\Omega_{\tilde{\chi}}h^2$ comes out too
large).

\par
As we already mentioned, near the $A$ pole, the
partial wave (or Taylor) expansion in
Eqs.~(\ref{taylorv}) and (\ref{average}) fails
\cite{coan,pole} badly. So, the thermal
averaging must by performed exactly using
numerical methods and employing the formulae in
Eqs.~(\ref{corraverage}), (\ref{pi2}), and
(\ref{corrsigma}). In order to achieve good
accuracy, it is also important to include the
one-loop QCD corrections \cite{width} to the
decay width of the $A$ particle entering its
propagator as well as to the quark masses.

\par
Another phenomenon which helps reducing
drastically $\Omega_{\tilde{\chi}}h^2$ and,
thus, satisfying the CDM constraint is strong
$\tilde{\chi}-\tilde{\tau}_2$ coannihilation
\cite{cdm,ellis1,ellis2} which operates when
$m_{\tilde{\tau}_2}$ gets close to
$m_{\tilde{\chi}}$. This yields
\cite{ellis1,ellis2} a relatively narrow
allowed region in the $m_0-M_{1/2}$ plane (for
fixed $A_0$ and $\tan\beta$), which stretches
just above the excluded region where the LSP is
the $\tilde{\tau}_2$.

\par
There exists \cite{funnel} also a ``bulk''
region at $m_0\sim M_{1/2}\sim {\rm few}\times
100~{\rm GeV}$ which is allowed by CDM
considerations. The (co)annihilation is
enhanced in this region due to the low values
of the various sparticle masses. However, this
region is, generally, excluded by other
phenomenological constraints (see
Sec.~\ref{sec:pheno}). So, the $A$-pole
annihilation of neutralinos and the
$\tilde{\chi}-\tilde{\tau}_2$ coannihilation
are the two basic available mechanisms for
obtaining acceptable values for the neutralino
relic abundance in the CMSSM.

\par
There are publicly available codes such as the
{\tt micrOMEGAs} \cite{micro} or the
{\tt DarkSUSY} \cite{darksusy} for the
calculation of $\Omega_{\tilde{\chi}}h^2$ in
MSSM which, among other improvements, include
all the relevant (co)annihilation channels
between all the sparticles (neutralinos,
charginos, squarks, sleptons, gluinos), use
exact tree-level cross sections, calculate
accurately and relativistically the thermal
averages, treat poles and final-state
thresholds properly, integrate the Boltzmann
equation numerically, and include the one-loop
QCD corrections to the decay widths of the
Higgs particles and the fermion masses. These
codes apply to any composition of the
neutralino and also include other
phenomenological constraints such as the
accelerator bounds on certain (s)particle
masses and the bounds on the anomalous magnetic
moment of the muon and the branching ration of
the process $b\rightarrow s\gamma$ (see
Sec.~\ref{sec:pheno}).

\section{Axinos}
\label{sec:axino}

\par
Another SUSY particle that could account for
the CDM in the universe is \cite{axinos} (see
also Ref.~\cite{leszek}) the axino $\tilde{a}$.
This particle, which is the superpartner of the
axion field, is a neutral Majorana chiral
fermion with negative R-parity. Its mass
$m_{\tilde{a}}$ is \cite{axinomass} strongly
model-dependent and can be anywhere in the
range $1~{\rm eV}-M_{\rm SUSY}$. In the limit
of unbroken SUSY, the axino mass is obviously
equal to the axion mass, which is tiny. Soft
SUSY breaking, however, generates suppressed
corrections to $m_{\tilde{a}}$ via
non-renormalizable operators of dimension five
or higher. So, the corrected mass is at most
of order $M_{\rm SUSY}^2/f_a\sim
1~{\rm keV}$ (note that no dimension-four soft
mass term is allowed for the axino since this
particle is a chiral fermion). In specific
SUSY models, there also exist one-loop
contributions to $m_{\tilde{a}}$, which are
typically $\stackrel{<}{_{\sim}}M_{\rm SUSY}$.
When the axion is a linear combination of the
phases of more than one superfields, we can
even have tree-level contributions to the
axino mass which can easily be as large as
$M_{\rm SUSY}$. In conclusion, $m_{\tilde{a}}$
is basically a free parameter ranging between
$1~{\rm eV}$ and $M_{\rm SUSY}$. This means
that the axino can easily be the LSP in SUSY
models.

\par
The axino couplings are suppressed by
$f_a$ with the most important of them being the
dimension-five axino ($\tilde{a}$)--gaugino
($\tilde{\lambda}$)--gauge boson ($A$)
Lagrangian coupling:
\begin{equation}
\mathcal{L}_{\tilde{a}\tilde{\lambda}A}=i\frac
{3\alpha_YC_{aYY}}{8\pi f_a}\bar{\tilde{a}}
\gamma_5[\gamma^\mu,\gamma^\nu]\tilde{B}
B_{\mu\nu}+i\frac{3\alpha_s}{8\pi f_a}
\bar{\tilde{a}}\gamma_5[\gamma^\mu,\gamma^\nu]
\tilde{g}^bF^b_{\mu\nu},
\label{axinocoupling}
\end{equation}
where $B$ and $\tilde{B}$ are, respectively,
the gauge boson and gaugino corresponding to
${\rm U}(1)_Y$, $F^b$ and $\tilde{g}^b$ the
gluon and gluino fields, $\alpha_Y$ and
$\alpha_s$ the ${\rm U}(1)_Y$ and strong
fine-structure constants, and $C_{aYY}$ a
model-dependent coefficient of order unity.

\par
Inflation dilutes utterly any pre-existing
axinos, which, after reheating, are {\em not}
in thermal equilibrium with the thermal bath
because of their very weak couplings
(suppressed by $f_a$). They can, however, be
thermally produced from the bath by 2-body
scattering processes or the decay of
(s)particles. The so-produced axinos are
initially relativistic, but out of thermal
equilibrium. This thermal production (TP) of
axinos is \cite{axinos} predominantly due to
2-body scattering processes of strongly
interacting particles (because of the relative
strength of strong interactions) involving the
$\tilde{a}\tilde{g}F$ coupling in
Eq.~(\ref{axinocoupling}). Such processes are
\begin{eqnarray}
g+g\rightarrow \tilde{a}+\tilde{g},\quad
g+\tilde{g}&\rightarrow&\tilde{a}+g,\quad
g+\tilde{q}\rightarrow\tilde{a}+q,\quad
g+q\rightarrow\tilde{a}+\tilde{q},
\nonumber \\
\tilde{q}+q\rightarrow\tilde{a}+g,\quad
\tilde{g}+\tilde{g}&\rightarrow&\tilde{a}+
\tilde{g},\quad
\tilde{g}+q\rightarrow\tilde{a}+q,\quad
\tilde{g}+\tilde{q}\rightarrow\tilde{a}+
\tilde{q},
\nonumber \\
q+\bar{q}&\rightarrow&\tilde{a}+\tilde{g},
\quad
\tilde{q}+\tilde{q}\rightarrow\tilde{a}+
\tilde{g},
\end{eqnarray}
where gluons and quarks are denoted by $g$ and
$q$ respectively. There exists \cite{axinos}
also TP of axinos from the decay of thermal
gluinos ($\tilde{g}\rightarrow\tilde{a}+g$) or
thermal neutralinos ($\tilde{\chi}\rightarrow
\tilde{a}+\gamma~[{\rm or}~Z]$). The latter
proceeds through the dimension-five Lagrangian
coupling $\tilde{a}\tilde{B}B$ in
Eq.~(\ref{axinocoupling}) provided
that the neutralino possesses an appreciable
bino component. These two decay processes are
important only for reheat temperatures
$T_{\rm r}$ of the order of the gluino mass
$m_{\tilde{g}}$ or the neutralino mass
$m_{\tilde{\chi}}$ respectively.

\par
There is also non-thermal production (NTP) of
axinos resulting from the decays of sparticles
which are out of thermal equilibrium. Indeed,
due to the suppressed couplings of the axino,
the sparticles first decay to the lightest
ordinary sparticle (LOSP), i.e. the lightest
sparticle with non-trivial SM quantum numbers,
which is the NLSP in this case. The LOSPs then
freeze out of thermal equilibrium and
eventually decay into axinos.

\par
\begin{figure}
\centering
\includegraphics[height=2.1in]{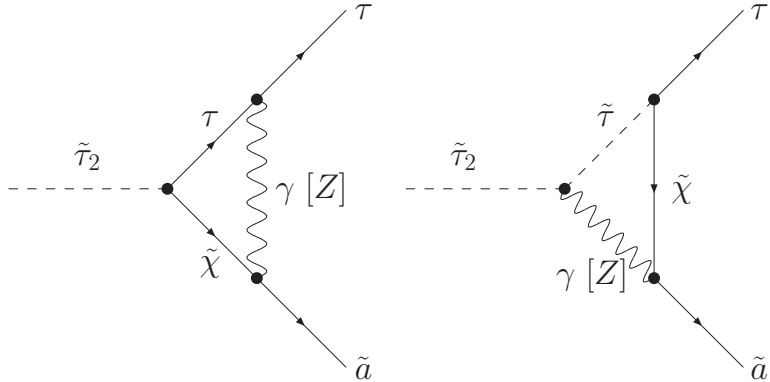}
\caption{The one-loop diagrams for the decay
$\tilde{\tau}_2\rightarrow\tau+\tilde{a}$.}
\label{fig:axino1}
\end{figure}

If the LOSP happens to be the lightest
neutralino, the relevant decay process is
\cite{axinos}
$\tilde{\chi}\rightarrow\tilde{a}+\gamma~
[{\rm or}~Z]$ through the coupling
$\tilde{a}\tilde{B}B$ in
Eq.~(\ref{axinocoupling})
provided that $\tilde{\chi}$ has a $\tilde{B}$
component. If, alternatively, the LOSP is the
lightest stau mass eigenstate, the decay
process for the NTP of axinos is \cite{small1}
$\tilde{\tau}_2 \rightarrow\tau+\tilde{a}$ via
the one-loop Feynman diagrams in
Fig.~\ref{fig:axino1}, which contain the
effective vertex $\tilde{\chi}\tilde{a}\gamma$
[or $\tilde{\chi}\tilde{a}Z$] from the coupling
$\tilde{a}\tilde{B}B$ in
Eq.~(\ref{axinocoupling}). In the decay of
$\tilde{\chi}$, $\gamma$'s and $q\bar{q}$ pairs
are produced. The latter originate from virtual
$\gamma$ and $Z$, or real $Z$ exchange and lead
to hadronic showers. In the $\tilde{\tau}_2$
case, the resulting $\tau$ decays immediately
into light mesons yielding again hadronic
showers. The electromagnetic and hadronic
showers emerging from the LOSP decay in both
cases, if they are generated after big bang
nucleosynthesis (BBN), can cause destruction
and/or overproduction of some of the light
elements, thereby jeopardizing the successful
predictions of BBN. This implies some
constraints on the parameters of the model
which, in the present case where the axino is
the LSP, come basically
from the hadronic showers alone due to the
relatively short LOSP lifetime.
In the case of a neutralino LOSP, we obtain
\cite{axinos} the bound $m_{\tilde{a}}
\stackrel{>}{_{\sim}}360~{\rm MeV}$ for low
values of the neutralino mass
$m_{\tilde{\chi}}$ ($\stackrel{<}{_{\sim}}60~
{\rm GeV}$), but no bound on the axino mass is
obtained for higher values of
$m_{\tilde{\chi}}$
($\stackrel{>}{_{\sim}}150~{\rm GeV}$).

\begin{figure}
\centering
\includegraphics[height=2.1in]{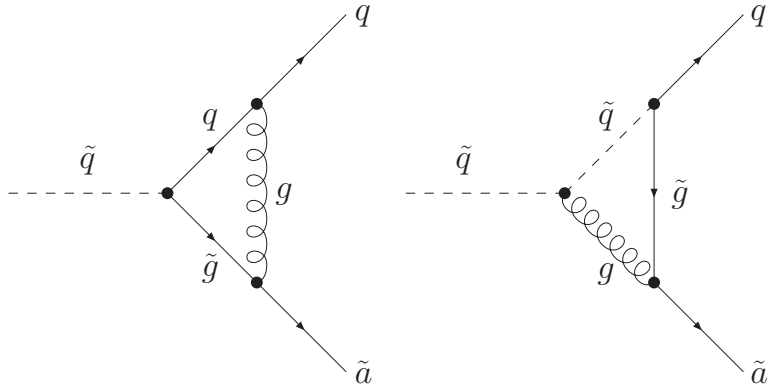}
\caption{The one-loop diagrams for the decay
$\tilde{q}\rightarrow q+\tilde{a}$.}
\label{fig:axino2}
\end{figure}

\par
\begin{figure}
\centering
\includegraphics[height=4.1in]{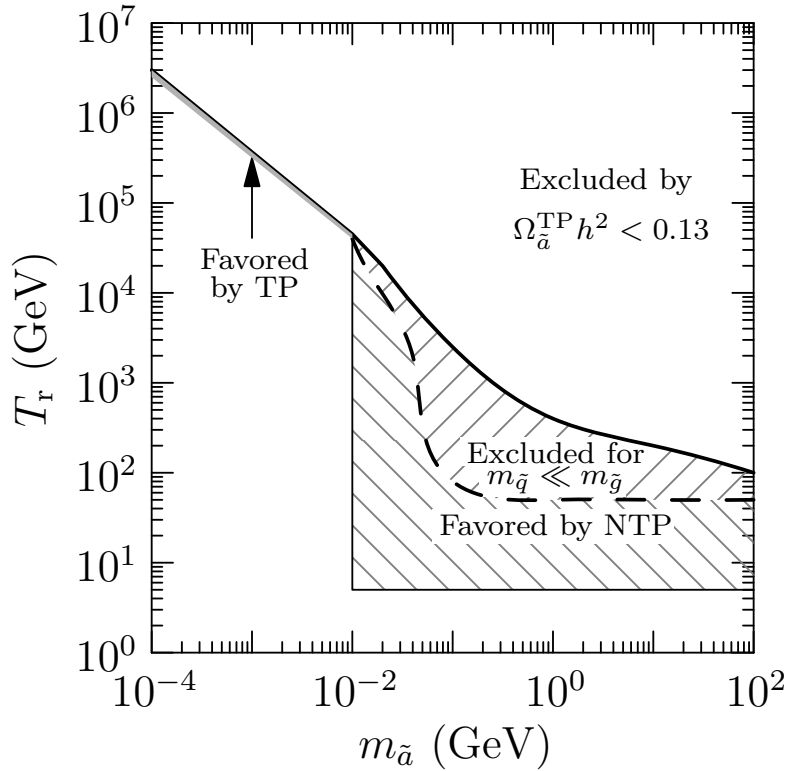}
\caption{The restrictions on the $m_{\tilde{a}}
-T_{\rm r}$ plane from axino CDM considerations
for $\tilde{\chi}=\tilde{B}$, $m_{\tilde{\chi}}
=100~{\rm GeV}$, $m_{\tilde{g}}=m_{\tilde{q}}=1
~{\rm TeV}$, and $f_a=10^{11}~{\rm GeV}$. The
solid almost diagonal line corresponds to
$\Omega_a^{\rm TP}h^2\approx 0.13$, where
$\Omega_a^{\rm TP}h^2$ is the TP axino
abundance. So, the area above this line is
cosmologically excluded. The narrow shaded area
just below the thin part of this line for
$m_{\tilde{a}}\stackrel{<}{_{\sim}}10~
{\rm MeV}$ is cosmologically favored by TP. The
hatched areas are favored by NTP. For
$m_{\tilde{q}}\ll m_{\tilde{g}}$,
the solid line is replaced by the dashed one,
whose position is strongly dependent on the
actual values of $m_{\tilde{q}}$,
$m_{\tilde{g}}$ and is only indicative here.
The area favored by NTP is then limited only to
the ``back-hatched'' region which lies below
the dashed line.}
\label{fig:axinoplot}
\end{figure}

We must further impose the following
constraints: ($a$) the predicted axino
abundance $\Omega_{\tilde{a}}h^2$ should lie in
the $95\%$ c.l. range for the CDM abundance in
the universe derived by the WMAP satellite
\cite{wmap}, ($b$) both the TP
and NTP axinos must become non-relativistic
before matter domination so as to contribute to
CDM, and ($c$) the NTP axinos should not
contribute too much relativistic energy density
during BBN since this can destroy its
successful predictions. For both $\tilde{\chi}$
or $\tilde{\tau}_2$ LOSP, the
requirements ($b$) and ($c$) imply that
$m_{\tilde{a}}\stackrel{>}{_{\sim}}100~
{\rm keV}$ or, equivalently, $T_{\rm r}
\stackrel{<}{_{\sim}}5\times 10^6~{\rm GeV}$.
For large values of the reheat temperature
($T_{\rm r}\stackrel{>}{_{\sim}}10^4~
{\rm GeV}$), TP of axinos is more efficient
than NTP and the cosmologically favored
region in parameter space where the
requirement ($a$) holds is quite narrow. For
smaller $T_{\rm r}$'s, NTP dominates yielding a
much wider favored region with $m_{\tilde{a}}
\stackrel{>}{_{\sim}}10~{\rm MeV}$. The upper
bound on $m_{\tilde{a}}$ increases as
$T_{\rm r}$ decreases towards
$m_{\tilde{\chi}}$. For $m_{\tilde{q}}\ll
m_{\tilde{g}}$, TP of axinos via the process
$\tilde{q}\rightarrow q+\tilde{a}$ becomes
\cite{small2} very
efficient leading to a reduction of the upper
limit on $T_{\rm r}$. As a result, the
cosmologically favored region from NTP is
reduced in this case. The Feynman diagrams for
the process $\tilde{q}\rightarrow q+\tilde{a}$
are depicted in Fig.~\ref{fig:axino2}. The
restrictions on the $m_{\tilde{a}}-T_{\rm r}$
plane from axino CDM considerations are
presented in Fig.~\ref{fig:axinoplot}.

\par
We find \cite{small1} that, for the CMSSM, with
appropriate choices of $m_{\tilde{a}}$ and
$T_{\rm r}$, almost any pair of values for
$m_0$ and $M_{1/2}$ can be allowed. This holds
for both $\tilde{\chi}$ or $\tilde{\tau}_2$ as
LOSP. However, the required $T_{\rm r}$'s for
achieving the WMAP bound on CDM turn out to be
quite low ($\stackrel{<}{_{\sim}}{\rm few}
\times 100~{\rm GeV}$).

\section{Gravitinos}
\label{sec:gravitino}

\par
It has been proposed
\cite{gravitino,gravitino1} that CDM could also
consist of gravitinos. The gravitino
$\tilde{G}$ is the superpartner of the graviton
and has negative R-parity. It can be the LSP in
many cases and, thus, contribute to CDM. In
the very CMSSM, its mass $m_{\tilde{G}}$ is
fixed by the asymptotic condition
$m_{3/2}=m_0$. In the general CMSSM, however,
it is a free parameter ranging between
$100~{\rm GeV}$ and $1~{\rm TeV}$. It can,
thus, very easily be the LSP in this case.

\par
The couplings of the gravitino are suppressed
by the Planck scale. The most important of them
are given by the dimension-five Lagrangian
terms
\begin{eqnarray}
\mathcal{L}&=&-\frac{1}{\sqrt{2}m_{\rm P}}
\mathcal{D}_\nu\phi^{i*}\bar{\tilde{\psi}}_\mu
\gamma^\nu\gamma^\mu\psi^i-\frac{1}{\sqrt{2}
m_{\rm P}}\mathcal{D}_\nu\phi^{i}
\bar{\psi}^i\gamma^\mu\gamma^\nu\tilde{\psi}_\mu
\nonumber \\
& &-\frac{i}{8m_{\rm P}}
\bar{\tilde{\psi}}_\mu[\gamma^\nu,\gamma^\rho]
\gamma^\mu\tilde{\lambda}^aF^a_{\nu\rho},
\label{gravitinocoupl}
\end{eqnarray}
where $\tilde{\psi}_\mu$ denotes the gravitino field,
$\phi^i$ are the complex scalar fields, $\psi^i$
are the corresponding chiral fermion fields,
$\tilde{\lambda}^a$ are the gaugino fields,
$m_{\rm P}\simeq 2.44\times 10^{18}~{\rm GeV}$
is the reduced Planck scale, and
$\mathcal{D}_\nu$ denotes the covariant
derivative. From these Lagrangian terms, we
obtain scalar--fermion--gravitino vertices
($\phi f\tilde{G}$) such as
$q\tilde{q}\tilde{G}$, $l\tilde{l}\tilde{G}$,
and $H\tilde{H}\tilde{G}$, as well as
gaugino--gauge boson--gravitino vertices
($\tilde{\lambda}F\tilde{G}$) such as
$g\tilde{g}\tilde{G}$ and $B\tilde{B}\tilde{G}$
(in this section, $l$ and $H$ represent any
lepton and Higgs boson respectively).

\par
The gravitinos are thermally produced after
reheating by $2\rightarrow 2$ scattering
processes involving the above vertices. Such
processes are \cite{gravitino,gravitino1}
\begin{eqnarray}
g+g\rightarrow\tilde{G}+\tilde{g},\quad
g+\tilde{g}&\rightarrow&\tilde{G}+g,\quad
g+\tilde{q}\rightarrow\tilde{G}+q,\quad
g+q\rightarrow\tilde{G}+\tilde{q},
\nonumber \\
q+\tilde{q}\rightarrow\tilde{G}+g,\quad
\tilde{g}+\tilde{g}&\rightarrow&
\tilde{G}+\tilde{g},\quad
\tilde{g}+q\rightarrow\tilde{G}+q,\quad
\tilde{g}+\tilde{q}\rightarrow\tilde{G}+
\tilde{q},
\nonumber \\
q+\bar{q}&\rightarrow&\tilde{G}+\tilde{g},\quad
\tilde{q}+\tilde{q}\rightarrow\tilde{G}+
\tilde{g}.
\end{eqnarray}

\par
There is \cite{gravitino1,gravitinoNTP} also
NTP of gravitinos via the decay of the
NLSP. For neutralino NLSP, the relevant decay
processes are $\tilde{\chi}\rightarrow\tilde{G}
+\gamma~[{\rm or}~Z]$ from the $\tilde{\lambda}
F\tilde{G}$ coupling and $\tilde{\chi}
\rightarrow\tilde{G}+H$ from the $H\tilde{H}
\tilde{G}$ coupling. In the case of
$\tilde{\tau}_2$ NLSP, the relevant decay
process is $\tilde{\tau}_2\rightarrow\tau+
\tilde{G}$ from the vertex $l\tilde{l}
\tilde{G}$. There is an important difference
between the NTP of gravitinos and axinos. In
the former case, the NLSP has a large lifetime
(up to about $10^8~{\rm sec}$). Consequently,
it gives rise mostly to electromagnetic, but
also to hadronic showers well after BBN. The
electromagnetic showers cause destruction of
some light elements (D, $^4{\rm He}$,
$^7{\rm Li}$) and/or overproduction of
$^3{\rm He}$ and $^6{\rm Li}$, thereby
disturbing BBN. The hadronic showers can also
disturb BBN. The overall resulting constraint
is \cite{gravitinoCmssm} very strong allowing
only limited regions of the parameter space of
the CMSSM lying exclusively in the range where
the NLSP is the $\tilde{\tau}_2$. Moreover, in
these allowed regions, the NTP of gravitinos is
not efficient enough to account for the
observed CDM abundance for
$M_{1/2}\stackrel{<}{_{\sim}}6~{\rm TeV}$.
However, we can compensate for the inefficiency
of NTP by raising $T_{\rm r}$ to enhance the TP
of $\tilde{G}$'s. The relic
gravitino abundance from TP, for
$m_{\tilde{G}}\ll m_{\tilde{g}}$, is
\cite{gravitinoTPformula}
\begin{equation}
\Omega^{\rm TP}_{\tilde{G}}h^2\approx 0.2\left(
\frac{T_{\rm r}}{10^{10}~{\rm GeV}}\right)
\left(\frac{100~{\rm GeV}}{m_{\tilde{G}}}
\right)\left(\frac{m_{\tilde{g}}(\mu)}
{1~{\rm TeV}}\right),
\end{equation}
where $m_{\tilde{g}}(\mu)$ is the running
gluino mass (for the general formula, see
Ref.~\cite{steffen}).

\section{Yukawa Quasi-Unification}
\label{sec:quasi}

\par
As already said in Sec.~\ref{sec:mssm},
exact YU in the framework of the CMSSM leads to
wrong values for $m_b$ and, thus, must be
corrected. We will now present a model which
naturally solves \cite{qcdm} (see also
Refs.~\cite{balkan,nova}) this $m_b$ problem
and discuss the restrictions on its parameter
space implied by CDM considerations and other
phenomenological constraints. Exact YU can be
achieved by embedding the MSSM in a SUSY GUT
model with a gauge group containing
${\rm SU(4)}_c$ and ${\rm SU(2)}_{\rm R}$.
Indeed, assuming that the electroweak Higgs
superfields $H_1$, $H_2$ and the third family
right handed quark superfields $t^c$, $b^c$
form ${\rm SU(2)}_{\rm R}$ doublets, we obtain
\cite{pana} the asymptotic Yukawa coupling
relation $h_t=h_b$ and, hence, large
$\tan\beta\sim m_{t}/m_{b}$. Moreover, if the
third generation quark and lepton
${\rm SU(2)}_{\rm L}$ doublets [singlets] $q_3$
and $l_3$ [$b^c$ and $\tau^c$] form a
${\rm SU(4)}_c$ {\bf 4}-plet
[${\bf\bar 4}$-plet] and the Higgs doublet
$H_1$ which couples to them is a
${\rm SU(4)}_c$ singlet, we get $h_b=h_{\tau}$
and the asymptotic relation $m_{b}=m_{\tau}$
follows. The simplest GUT gauge group which
contains both ${\rm SU(4)}_c$ and
${\rm SU(2)}_{\rm R}$ is the Pati-Salam (PS)
group $G_{\rm PS}={\rm SU(4)}_c\times
{\rm SU(2)}_{\rm L}\times {\rm SU(2)}_{\rm R}$
and we will use it here.

\par
As mentioned, applying YU in the context of the
CMSSM and given the experimental values of the
top-quark and tau-lepton masses (which
naturally restrict $\tan\beta\sim 50$), the
resulting value of the $b$-quark mass turns out
to be unacceptable. This is due to the fact
that, in the large $\tan\beta$ regime, the
tree-level $b$-quark mass receives sizeable
SUSY corrections \cite{pierce,copw,susy,king}
(about 20$\%$), which have the sign of $\mu$
(with the standard sign convention
\cite{sugra}) and drive, for $\mu>[<]~0$, the
corrected $b$-quark mass at $M_Z$, $m_b(M_Z)$,
well above [somewhat below] its $95\%$ c.l.
experimental range
\begin{equation}
2.684~{\rm GeV}\stackrel{<}{_{\sim}}
m_b(M_Z)
\stackrel{<}{_{\sim}} 3.092~{\rm
GeV}\quad\mbox{with}\quad\alpha_s(M_Z)=0.1185.
\label{mbrg}
\end{equation}
This is derived by appropriately \cite{qcdm}
evolving the corresponding range of $m_b(m_b)$
in the $\overline{MS}$ scheme (i.e.
$3.95-4.55~{\rm GeV}$) up to $M_{Z}$ in
accordance with Ref.~\cite{baermb}. We see
that, for both signs of $\mu$, YU leads to an
unacceptable $b$-quark mass with the $\mu<0$
case being less disfavored.

\par
A way out of this $m_b$ problem can be found
\cite{qcdm} (see also Refs.~\cite{balkan,nova})
without having to abandon the CMSSM (in
contrast to
the usual strategy \cite{king,raby,baery,nath})
or YU altogether. We can rather modestly
correct YU by including an extra
${\rm SU(4)}_c$ non-singlet Higgs superfield
with Yukawa couplings to the quarks and
leptons. The Higgs ${\rm SU(2)}_{\rm L}$
doublets
contained in this superfield can naturally
develop \cite{wetterich} subdominant VEVs and
mix with the main electroweak doublets, which
are assumed to be ${\rm SU(4)}_c$ singlets and
form a ${\rm SU(2)}_{\rm R}$ doublet. This
mixing can, in general, violate the
${\rm SU(2)}_{\rm R}$ symmetry.
Consequently, the resulting electroweak Higgs
doublets $H_1$, $H_2$ do not form a
${\rm SU(2)}_{\rm R}$ doublet and also break the
${\rm SU}(4)_c$ symmetry. The required deviation
from YU is expected to be more pronounced for
$\mu>0$. Despite this, we will study here this
case, since the $\mu<0$ case has been excluded
\cite{cd2} by combining the WMAP restrictions
\cite{wmap} on the CDM in the universe with the
experimental results \cite{cleo} on the
inclusive branching ratio
${\rm BR}(b\rightarrow s\gamma)$. The same SUSY
GUT model which, for $\mu>0$ and universal
boundary conditions, remedies the $m_b$ problem
leads to a new version \cite{jean2} of shifted
hybrid inflation \cite{jean}, which, as the
older version \cite{jean}, avoids monopole
overproduction at the end of inflation,
but, in contrast to that version, is based only
on renormalizable interactions.

\par
In Sec.~\ref{sec:model}, we review the
construction of a SUSY GUT model which
naturally and modestly violates YU, yielding an
appropriate Yukawa quasi-unification condition
(YQUC), which is derived in Sec.~\ref{sec:yquc}.
We then outline the resulting CMSSM in
Sec.~\ref{sec:qcmssm} and introduce the various
cosmological and phenomenological requirements
which restrict its parameter space in
Sec.~\ref{sec:pheno}. In
Sec.~\ref{sec:parameters}, we delineate the
allowed range of parameters. Finally, in
Sec.~\ref{sec:shift}, we briefly comment on the
new version of shifted hybrid inflation which
is realized in this model.

\subsection{The PS SUSY GUT Model}
\label{sec:model}

We will take the SUSY GUT model of shifted
hybrid inflation \cite{jean} (see also
Ref.~\cite{talks}) as our starting point. It is
based on $G_{\rm PS}$, which is the simplest
GUT gauge group that can lead to exact YU. The
representations under $G_{\rm PS}$ and the
global charges of the various matter and Higgs
superfields contained in this model are
presented in Table~\ref{tab:charges}, which
also contains the extra Higgs superfields
required for accommodating an adequate
violation of YU for $\mu>0$ (see below). The
matter superfields are $F_i$ and $F^c_i$
($i=1,2,3$), while the electroweak Higgs
doublets belong to the superfield $h$. So, all
the requirements for exact YU are fulfilled.
The spontaneous breaking of
$G_{\rm PS}$ down to $G_{\rm SM}$ is achieved
by the superheavy VEVs ($\sim M_{\rm GUT}$) of
the right handed neutrino-type components of a
conjugate pair of Higgs superfields $H^c$,
$\bar{H}^c$. The model also contains a gauge
singlet $S$ which
triggers the breaking of $G_{\rm PS}$, a
${\rm SU(4)}_c$ {\bf 6}-plet $G$ which gives
\cite{leontaris} masses to the right handed
down-quark-type components of $H^c$,
$\bar{H}^c$, and a pair of gauge singlets $N$,
$\bar{N}$ for solving \cite{rsym} the $\mu$
problem of the MSSM via a PQ symmetry (for an
alternative solution of the $\mu$ problem, see
Ref.~\cite{dvali}).
In addition to $G_{\rm PS}$, the model possesses
two global ${\rm U(1)}$ symmetries, namely a R
and a PQ symmetry, as well as the discrete
matter parity symmetry $Z_2^{\rm mp}$. Note
that global continuous symmetries such as our
PQ and R symmetry can effectively arise
\cite{laz1} from the rich discrete symmetry
groups encountered in many compactified string
theories (see e.g. Ref.~\cite{laz2}). Note
that, although the model contains baryon- and
lepton-number-violating superpotential terms,
the proton is \cite{qcdm,jean} practically
stable. The baryon asymmetry of the universe
is generated via the non-thermal realization
\cite{origin} of the leptogenesis scenario
\cite{lepto} (for recent papers on non-thermal
leptogenesis, see e.g. Ref.~\cite{leptosc}).

\begin{table}
\centering
\caption{Superfield Content of the Model}
{\footnotesize
\begin{tabular}{|ccccc|}
\hline {} &{} &{} &{} &{} \\ [-2.2ex]
{~~~Superfields~~~}&{~~~Representations~~~}&
\multicolumn{3}{c|}{Global}
\\ [0.4ex]
\multicolumn{1}{|c}{}&{under $G_{\rm PS}$}
&\multicolumn{3}{c|}{~~~Charges~~~}
\\ [0.4ex]
{}&{}&{$R$}&{$PQ$}&{$Z^{\rm mp}_2$}
\\ [0.4ex]
\hline {} &{} &{} &{} &{} \\ [-2.2ex]
\multicolumn{5}{|c|}{Matter Superfields}
\\ [0.4ex]
\hline {} &{} &{} &{} &{} \\ [-2.2ex]
{$F_i$} &{$({\bf 4, 2, 1})$}& $1/2$ & $-1$ &$1$
\\ [0.4ex]
{$F^c_i$} & {$({\bf \bar 4, 1, 2})$} &{$1/2$}
&{$0$}&{$-1$}
\\ [0.4ex]
\hline {} &{} &{} &{} &{} \\ [-2.2ex]
\multicolumn{5}{|c|}{Higgs Superfields}
\\ [0.4ex]
\hline {} &{} &{} &{} &{}\\ [-2.2ex]
{$h$} & {$({\bf 1, 2, 2})$}&$0$ &$1$ &$0$
\\ [0.4ex]
\hline {} &{} &{} &{} &{} \\ [-2.2ex]
{$H^c$}&{$({\bf\bar 4,1,2})$}&{$0$}&{$0$}&{$0$}
\\ [0.4ex]
{$\bar H^c$}&$({\bf 4, 1, 2})$&{$0$}&{$0$}&
{$0$}
\\ [0.4ex]
{$S$} & {$({\bf 1,1,1})$}&$1$ &$0$ &$0$
\\ [0.4ex]
{$G$} & {$({\bf 6,1,1})$}&$1$ &$0$ &$0$
\\ [0.4ex]
\hline {} & {} & {} & {} & {} \\ [-2.2ex]
{$N$} &{$({\bf 1, 1, 1})$}&{$1/2$}&{$-1$}&{$0$}
\\ [0.4ex]
{$\bar N$}&$({\bf 1, 1, 1})$& {$0$}&{$1$}&{$0$}
\\ [0.4ex]
\hline {} &{} &{} &{} &{} \\ [-2.2ex]
\multicolumn{5}{|c|}{Extra Higgs Superfields}
\\ [0.4ex]
\hline {} &{} &{} &{} &{} \\ [-2.2ex]
$h^{\prime}$&{$({\bf 15,2,2})$}&$0$&$1$&$0$
\\ [0.4ex]
$\bar h^{\prime}$&{$({\bf 15,2,2})$}&$1$&$-1$
&$0$
\\ [0.4ex]
$\phi$&$({\bf 15, 1, 3})$ & $0$ & $0$ &$0$
\\ [0.4ex]
$\bar\phi$&{$({\bf 15, 1, 3})$}&$1$&$0$&$0$
\\ [0.4ex]
\hline
\end{tabular}}
\label{tab:charges}
\end{table}

\par
A moderate violation of exact YU can be
naturally accommodated in this model by adding
a new Higgs superfield $h^{\prime}$ with Yukawa
couplings $FF^ch^{\prime}$. Actually,
({\bf 15,2,2}) is the only representation of
$G_{\rm PS}$, besides ({\bf 1,2,2}), which
possesses such couplings to the matter
superfields. In order to give superheavy masses
to the color non-singlet components of
$h^{\prime}$, we need to include one more Higgs
superfield $\bar{h}^{\prime}$ with the
superpotential coupling
$\bar{h}^{\prime}h^{\prime}$, whose coefficient
is of the order of $M_{\rm GUT}$.

\par
After the breaking of $G_{\rm PS}$ to
$G_{\rm SM}$, the two color singlet
${\rm SU(2)}_{\rm L}$ doublets $h_1^{\prime}$,
$h_2^{\prime}$ contained in $h^{\prime}$ can
mix with the corresponding doublets $h_1$,
$h_2$ in $h$. This is mainly due to the terms
$\bar{h}^{\prime}h^{\prime}$ and
$H^c\bar{H}^c\bar{h}^{\prime}h$. Actually,
since
\begin{eqnarray}
&& H^c\bar H^c=({\bf\bar 4,1,2})({\bf 4,1,2})=
({\bf 15,1,1+3})+\cdots,
\nonumber \\
&& \bar h^\prime h=({\bf 15,2,2})({\bf 1,2,2})=
({\bf 15, 1, 1+3})+\cdots,
\end{eqnarray}
there are two independent couplings of the type
$H^c\bar{H}^c\bar{h}^{\prime}h$ (both
suppressed by the string scale
$M_{\rm S}\approx 5\times 10^{17}~{\rm GeV}$,
as they are non-renormalizable). One of these
couplings is
between the ${\rm SU(2)}_{\rm R}$ singlets in
$H^c\bar{H}^c$ and $\bar{h}^{\prime}h$ and the
other between the ${\rm SU(2)}_{\rm R}$
triplets in these combinations. So, we obtain
two bilinear terms $\bar{h}_1^{\prime}h_1$ and
$\bar{h}_2^{\prime}h_2$ with different
coefficients, which are suppressed by
$M_{\rm GUT}/M_{\rm S}$. These terms together
with the terms $\bar{h}_1^{\prime}h_1^{\prime}$
and $\bar{h}_2^{\prime}h_2^{\prime}$ from
$\bar{h}^{\prime}h^{\prime}$, which have equal
coefficients, generate different mixings
between $h_1$, $h_1^{\prime}$ and $h_2$,
$h_2^{\prime}$. Consequently, the resulting
electroweak doublets $H_1$, $H_2$ contain
${\rm SU}(4)_c$ violating components suppressed
by $M_{\rm GUT}/M_{\rm S}$ and fail to form a
${\rm SU(2)}_{\rm R}$ doublet by an equally
suppressed amount. So, YU is naturally and
moderately violated. Unfortunately, as it turns
out, this violation is not adequately large for
correcting the bottom-quark mass within the
framework of the CMSSM with $\mu>0$.

\par
In order to allow for a more sizable violation
of YU, we further extend the model by including
the superfield $\phi$ with the coupling
$\phi\bar{h}^{\prime}h$. To give superheavy
masses to the color non-singlets in $\phi$, we
introduce one more superfield $\bar\phi$ with
the coupling $\bar\phi\phi$, whose coefficient
is of order $M_{\rm GUT}$.

\par
The superpotential terms $\bar\phi\phi$ and
$\bar\phi H^c\bar{H}^c$ imply that, after the
breaking of $G_{\rm PS}$ to $G_{\rm SM}$,
$\phi$ acquires a VEV of order $M_{\rm GUT}$.
The coupling $\phi\bar{h}^{\prime}h$ then
generates ${\rm SU(2)}_{\rm R}$ violating
unsuppressed bilinear terms between the
doublets in $\bar{h}^{\prime}$ and $h$. These
terms can overshadow the corresponding ones
from the non-renormalizable term
$H^c\bar{H}^c\bar{h}^{\prime}h$. The resulting
${\rm SU(2)}_{\rm R}$ violating mixing of the
doublets in $h$ and $h^{\prime}$ is then
unsuppressed and we can obtain stronger
violation of YU.

\subsection{The YQUC}
\label{sec:yquc}

\par
To further analyze the mixing of the doublets
in $h$ and $h^{\prime}$, observe that the part
of the superpotential corresponding to the
symbolic couplings
$\bar{h}^{\prime}h^{\prime}$,
$\phi\bar{h}^{\prime}h$ is properly written as
\begin{equation}
m{\rm tr}\left(\bar h^{\prime}\epsilon
h^{\prime\,{\rm T}}\epsilon\right)+p{\rm tr}
\left(\bar h^{\prime}
\epsilon\phi h^{\rm T}\epsilon\right),
\label{expmix}
\end{equation}
where $m$ is a mass parameter of order
$M_{\rm GUT}$, $p$ is a dimensionless parameter
of order unity, ${\rm tr}$ denotes trace taken
with respect to the ${\rm SU(4)}_{\rm c}$ and
${\rm SU(2)}_{\rm L}$ indices, and the
superscript T denotes the transpose of a
matrix.

\par
After the breaking of $G_{\rm PS}$ to
$G_{\rm SM}$, $\phi$ acquires a VEV
$\langle\phi\rangle\sim M_{\rm GUT}$.
Substituting it by this VEV in the above
couplings, we obtain
\begin{eqnarray}
&& {\rm tr}(\bar{h}^{\prime}\epsilon
h^{\prime\,{\rm T}}\epsilon)=
\bar{h}^{\prime\,{\rm T}}_1\epsilon
h^{\prime}_2+h^{\prime\,{\rm T}}_1\epsilon
\bar{h}^{\prime}_2+\cdots,
\label{mass}
\\
&& {\rm tr}(\bar{h}^{\prime}\epsilon
\phi h^{\rm T}\epsilon)=
\frac{\langle\phi\rangle}{\sqrt{2}}{\rm tr}
(\bar{h}^{\prime}\epsilon\sigma_3h^{\rm T}
\epsilon)=\frac{\langle\phi\rangle}{\sqrt{2}}
(\bar{h}^{\prime\,{\rm T}}_1\epsilon h_2-
h_1^{\rm T}\epsilon\bar{h}^{\prime}_2),
\label{triplet}
\end{eqnarray}
where the ellipsis in Eq.~(\ref{mass}) contains
the colored components of $\bar{h}^{\prime}$,
$h^{\prime}$ and $\sigma_3={\rm diag}(1,-1)$.
Inserting Eqs.~(\ref{mass}) and (\ref{triplet})
into Eq.~(\ref{expmix}), we obtain
\begin{equation}
m\bar h^{\prime\,{\rm T}}_1\epsilon(h^{\prime}_2-
\alpha h_2)+m(h^{\prime\,{\rm T}}_1+\alpha
h^{\rm T}_1)\epsilon\bar{h}^{\prime}_2\quad
\mbox{with}\quad\alpha=-\frac{p\langle\phi
\rangle}{\sqrt{2}m}.
\label{superheavy}
\end{equation}
So, we get two pairs of superheavy doublets
with mass $m$. They are predominantly given by
\begin{equation}
\bar{h}^{\prime}_1,~\frac{h^{\prime}_2-
\alpha h_2}{\sqrt{1+|\alpha|^2}}\quad
{\rm and}\quad\frac{h^{\prime}_1+\alpha h_1}
{\sqrt{1+|\alpha|^2}},~\bar{h}^{\prime}_2.
\label{superdoublets}
\end{equation}
The orthogonal combinations of $h_1$,
$h^{\prime}_1$ and $h_2$, $h^{\prime}_2$
constitute the electroweak doublets
\begin{equation}
H_1=\frac{h_1-\alpha^*h^{\prime}_1}
{\sqrt{1+|\alpha|^2}}\quad{\rm and}\quad
H_2=\frac{h_2+\alpha^*h^{\prime}_2}
{\sqrt{1+|\alpha|^2}}.
\label{ew}
\end{equation}
The superheavy doublets in
Eq.~(\ref{superdoublets}) must have vanishing
VEVs, which readily implies that
$\langle h_1^{\prime}\rangle=-\alpha\langle
h_1\rangle$ and $\langle h_2^{\prime}\rangle=
\alpha\langle h_2\rangle$.
Equation~(\ref{ew}) then gives
$\langle H_1\rangle=(1+|\alpha|^2)^{1/2}
\langle h_1 \rangle$, $\langle H_2\rangle=
(1+|\alpha|^2) ^{1/2}\langle h_2\rangle$.
From the third generation Yukawa couplings
$y_{33}F_3hF_3^c$,
$2y_{33}^{\prime}F_3h^{\prime}F_3^c$, we obtain
\begin{eqnarray}
&&m_t=|y_{33}\langle h_2\rangle+y_{33}^{\prime}
\langle
h_2^{\prime}\rangle|=\left|\frac{1+\rho\alpha
/\sqrt{3}}{\sqrt{1+|\alpha|^2}}y_{33} \langle
H_2\rangle\right|,\label{top} \\ &&
m_b=\left|\frac{1-\rho\alpha/
\sqrt{3}}{\sqrt{1+|\alpha|^2}}y_{33} \langle
H_1\rangle\right|,~m_\tau=
\left|\frac{1+\sqrt{3}\rho\alpha}
{\sqrt{1+|\alpha|^2}}y_{33} \langle
H_1\rangle\right|,
\label{bottomtau}
\end{eqnarray}
where $\rho=y_{33}^{\prime}/y_{33}$. From
Eqs.~(\ref{top}) and (\ref{bottomtau}), we see
that YU is now replaced by the YQUC
\begin{equation}
h_t:h_b:h_\tau=(1+c):(1-c):(1+3c)\quad
\mbox{with}\quad
0<c=\rho\alpha/\sqrt{3}<1.
\label{minimal}
\end{equation}
For simplicity, we restricted ourselves here to
real values of $c$ only which lie between zero
and unity, although $c$ is, in general, an
arbitrary complex quantity with $|c|\sim 1$.

\subsection{The Resulting CMSSM}
\label{sec:qcmssm}

Below the GUT scale $M_{\rm GUT}$, the particle
content of our model reduces to this of MSSM
(modulo SM singlets). We assume universal soft
SUSY breaking scalar masses $m_0$, gaugino
masses $M_{1/2}$, and trilinear scalar
couplings $A_0$ at $M_{\rm GUT}$. Therefore,
the resulting MSSM is the so-called CMSSM
\cite{Cmssm} with $\mu>0$ supplemented by the
YQUC in Eq.~(\ref{minimal}). With these initial
conditions, we run the MSSM RGEs \cite{cdm}
between $M_{\rm GUT}$ and a common variable
SUSY threshold $M_{\rm SUSY}$ (see
Sec.~\ref{sec:mssm}) determined in consistency
with the SUSY spectrum of the model. At
$M_{\rm SUSY}$, we impose radiative electroweak
symmetry breaking, evaluate the SUSY spectrum
and incorporate the SUSY corrections
\cite{pierce,susy,king} to the $b$-quark and
$\tau$-lepton masses. Note that the corrections
to the $\tau$-lepton mass (almost 4$\%$) lead
\cite{cd2} to a small reduction of $\tan\beta$.
From $M_{\rm SUSY}$ to $M_Z$, the running of
gauge and Yukawa coupling constants is
continued using the SM RGEs.

\par
For presentation purposes, $M_{1/2}$ and $m_0$
can be replaced \cite{cdm} by the LSP mass
$m_{\rm LSP}$ and the relative mass splitting
between this particle and the lightest stau
$\Delta_{\tilde\tau_2}=(m_{\tilde\tau_2}
-m_{\rm LSP })/m_{\rm LSP}$ (recall that
$\tilde{\tau}_2$ is the NLSP in this case). For
simplicity, we restrict this presentation to
the $A_0=0$ case (for $A_0\neq0$ see
Refs.~\cite{qcdm,mario}). So, our input
parameters are $m_{\rm LSP}$,
$\Delta_{\tilde\tau_2}$, $c$, and $\tan\beta$.

\par
For any given $m_b(M_Z)$ in the range in
Eq.~(\ref{mbrg}) and with fixed
$m_t(m_t)=166~{\rm GeV}$ and
$m_\tau(M_Z)=1.746~{\rm GeV}$, we can
determine the parameters $c$ and $\tan\beta$ at
$M_{\rm SUSY}$ so that the YQUC in
Eq.~(\ref{minimal}) is satisfied. We are, thus,
left with $m_{\rm LSP}$ and
$\Delta_{\tilde\tau_2}$ as free parameters.

\subsection{Cosmological and Phenomenological
Constraints}
\label{sec:pheno}

\par
Restrictions on the parameters of our model can
be derived by imposing a number of cosmological
and phenomenological requirements (for similar
recent analyses,
see Refs.~\cite{baery,nath,spanos}). These
constraints result from

$\bullet$ {\em CDM Considerations}. As
discussed in Sec.~\ref{sec:mssm}, in the
context of the CMSSM, the LSP can be the
lightest neutralino which is an almost pure
bino. It naturally arises \cite{goldberg}
as a CDM candidate. We require its relic
abundance, $\Omega_{\rm LSP}h^2$, not to exceed
the $95\%$ c.l. upper bound on the CDM
abundance derived \cite{wmap} by WMAP:
\begin{equation}
\Omega_{\rm CDM}h^2\stackrel{<}{_{\sim}} 0.13.
\label{cdmb}
\end{equation}
We calculate $\Omega_{\rm LSP}h^2$ using {\tt
micrOMEGAs} \cite{micro}, which is certainly
one of the most complete publicly available
codes. Among other things, it includes all
possible coannihilation processes \cite{ellis2}
and one-loop QCD corrections \cite{width} to
the Higgs decay widths and couplings to
fermions.

$\bullet$ {\em Branching Ratio of
$b\rightarrow s\gamma$}. Taking into account
the experimental results of Ref.~\cite{cleo} on
this ratio, ${\rm BR}(b\rightarrow s\gamma)$,
and combining \cite{qcdm} appropriately the
experimental and theoretical errors involved,
we obtain the $95\%$ c.l. range
\begin{equation}
1.9\times 10^{-4}\stackrel{<}{_{\sim}}
{\rm BR}(b\rightarrow s\gamma)
\stackrel{<}{_{\sim}} 4.6 \times 10^{-4}.
\label{bsgb}
\end{equation}
Although there exist more recent
experimental data \cite{babar} on the branching
ratio of $b\rightarrow s\gamma$, we do not use
them here. The reason is that these data do not
separate the theoretical errors from the
experimental ones and, thus, the derivation of
the $95\%$ c.l. range is quite ambiguous. In
any case, the $95\%$ c.l. limits obtained in
Ref.~\cite{babarellis} on the basis of these
latest measurements are not terribly different
from the ones quoted in Eq.~(\ref{bsgb}). In
view of this and the fact that, in our case,
the restrictions from
${\rm BR}(b\rightarrow s\gamma)$ are
overshadowed by other constraints (see
Sec.~\ref{sec:parameters}), we limit ourselves
to the older data. We compute
${\rm BR}(b\rightarrow s\gamma)$ by using an
updated version of the relevant calculation
contained in the {\tt micrOMEGAs} package
\cite{micro}. In this code, the SM contribution
is calculated following Ref.~\cite{kagan}. The
charged Higgs ($H^\pm$) contribution is
evaluated by including the next-to-leading
order (NLO) QCD corrections \cite{nlo} and
$\tan\beta$ enhanced contributions \cite{nlo}.
The dominant SUSY contribution includes
resummed NLO SUSY QCD corrections \cite{nlo},
which hold for large $\tan\beta$.

$\bullet$ {\em Muon Anomalous Magnetic Moment}.
The deviation, $\delta a_\mu$, of the measured
value of $a_\mu$ from its predicted value in
the SM, $a^{\rm SM}_\mu$, can be attributed to
SUSY contributions, which are calculated by
using the {\tt micrOMEGAs} routine
\cite{gmuon}. The
calculation of $a^{\rm SM}_\mu$ is not yet
stabilized mainly because of the instability of
the hadronic vacuum polarization contribution.
According to recent calculations (see e.g.
Refs.~\cite{davier,vainshtein}), there is still
a considerable discrepancy between the findings
based on the $e^+e^-$ annihilation data and the
ones based on the $\tau$-decay data. Taking
into account the results of Ref.~\cite{davier}
and the experimental measurement of $a_\mu$
reported in Ref.~\cite{muon}, we get the
following $95\%$ c.l. ranges:
\begin{eqnarray}
-0.53\times10^{-10}\stackrel{<}{_{\sim}}&\delta
a_\mu&\stackrel{<}{_{\sim}} 44.7\times
10^{-10},~\quad\mbox{$e^+e^-$-based};
\label{g2e}
\\ \vspace*{19pt}
-13.6\times10^{-10}\stackrel{<}{_{\sim}}&
\delta a_\mu&\stackrel{<}{_{\sim}} 28.4
\times 10^{-10},\quad\mbox{$\tau$-based}.
\label{g2t}
\end{eqnarray}
Following the common practice \cite{spanos}, we
adopt the restrictions to parameters induced by
Eq.~(\ref{g2e}), since Eq.~(\ref{g2t}) is
considered as quite oracular, due to poor
$\tau$-decay data. It is true that there exist
more recent experimental data \cite{muonn} on
$a_\mu$ than the ones we considered and more
updated estimates of $\delta a_\mu$ than the
one in Ref.~\cite{davier} (see e.g.
Ref.~\cite{vainshtein}). However, only the
$95\%$ c.l. upper limit on $\delta a_\mu$
enters into our analysis here and its new
values are not very different from the one in
Eq.~(\ref{g2e}).

$\bullet$ {\em Collider Bounds}. Here, as it
turns out, the only relevant collider bound is
the $95\%$ c.l. LEP lower bound \cite{higgs} on
the mass of the lightest $CP$-even neutral
Higgs boson $h$:
\begin{equation}
m_h\stackrel{>}{_{\sim}} 114.4~{\rm GeV}.
\label{mhb}
\end{equation}
The SUSY corrections to the lightest $CP$-even
Higgs boson mass $m_h$ are calculated at two
loops by using the {\tt FeynHiggsFast}
program \cite{fh} included in the
{\tt micrOMEGAs} code \cite{micro}.

\subsection{The Allowed Parameter Space}
\label{sec:parameters}

\par
We will now try to delineate the parameter
space of our model with $\mu>0$ which is
consistent with the constraints in
Sec.~\ref{sec:pheno}. The restrictions on the
$m_{\rm LSP}-\Delta_{\tilde\tau_2}$ plane, for
$A_0=0$ and the central values of
$\alpha_s(M_Z)=0.1185$ and
$m_b(M_Z)=2.888~{\rm GeV}$, are indicated in
Fig.~\ref{figa} by solid lines, while the upper
bound on $m_{\rm LSP}$ from Eq.~(\ref{cdmb}),
for $m_b(M_Z)=2.684~[3.092]~{\rm GeV}$, is
depicted by a dashed [dotted] line. We observe
the following:

\begin{figure}
\centering
\includegraphics[height=4.1in,angle=-90]
{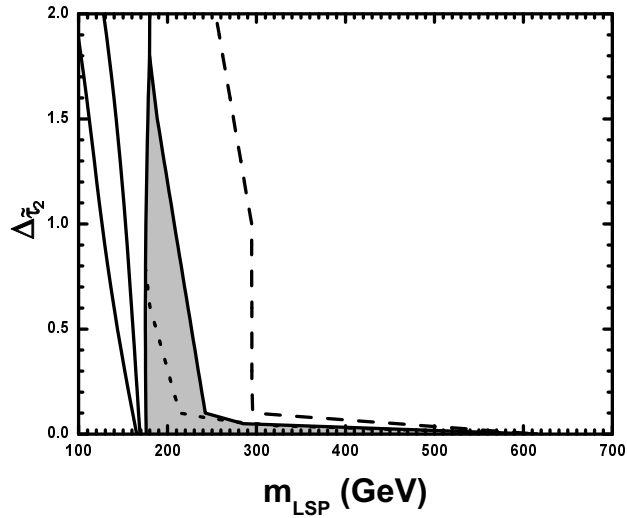}
\caption{The various restrictions on the
$m_{\rm LSP}-\Delta_{\tilde\tau_2}$ plane for
$\mu>0$, $A_0=0$, and $\alpha_s(M_Z)=0.1185$.
From left to right, the solid lines depict the
lower bounds on $m_{\rm LSP}$ from
$\delta a_\mu<44.7\times 10^{-10}$, ${\rm BR}
(b\rightarrow s\gamma)>1.9\times 10^{-4}$, and
$m_h>114.4~{\rm GeV}$ and the upper bound on
$m_{\rm LSP}$ from $\Omega_{\rm LSP}h^2<0.13$
for $m_b(M_Z)=2.888~{\rm GeV}$. The dashed
[dotted] line depicts the upper bound on
$m_{\rm LSP}$ from $\Omega_{\rm LSP}h^2<0.13$ for
$m_b(M_Z)=2.684~[3.092]~{\rm GeV}$. The allowed
area for $m_b(M_Z)=2.888~{\rm GeV}$ is shaded.}
\label{figa}
\end{figure}

\begin{figure}
\centering
\includegraphics[height=4.1in,angle=-90]
{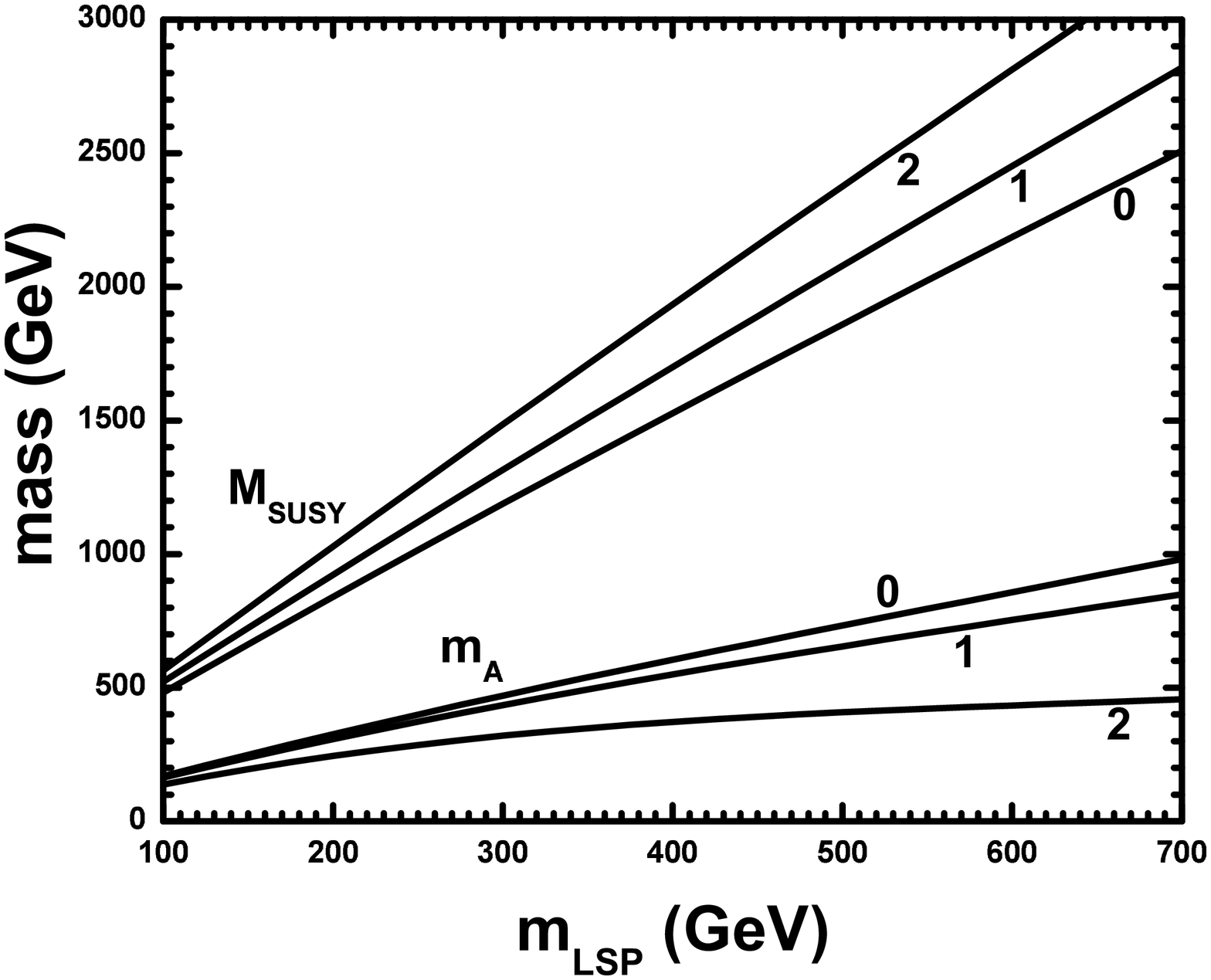}
\caption{The mass parameters $m_A$ and
$M_{\rm SUSY}$ versus $m_{\rm LSP}$
for various values of $\Delta_{\tilde\tau_2}$,
which are indicated on the curves. We take
$\mu>0$, $A_0=0$, $m_b(M_Z)=2.888~{\rm GeV}$,
and $\alpha_s(M_Z)=0.1185$.}
\label{figb}
\end{figure}

\begin{itemize}

\item The lower bounds on $m_{\rm LSP}$ are not
so sensitive to the variations of $m_b(M_Z)$.

\item The lower bound on $m_{\rm LSP}$ from
Eq.~(\ref{mhb}) overshadows all the other lower
bounds on this mass.

\item The upper bound on $m_{\rm LSP}$ from
Eq.~(\ref{cdmb}) is very sensitive to the
variations of $m_b(M_Z)$. In particular, one
notices the extreme sensitivity of the almost
vertical part of the corresponding line, where
the LSP annihilation via an $A$-boson exchange
in the s-channel is \cite{lah} by far the
dominant process, since $m_A$, which is smaller
than $2m_{\rm LSP}$, is always very close to it
as seen from Fig.~\ref{figb}. This sensitivity
can be understood from Fig.~\ref{figc}, where
$m_A$ is depicted versus $m_{\rm LSP}$ for
various $m_b(M_Z)$'s. We see that, as
$m_b(M_Z)$ decreases, $m_A$ increases and
approaches $2m_{\rm LSP}$. The $A$-pole
annihilation is then enhanced and
$\Omega_{\rm LSP}h^2$ is drastically reduced
causing an increase of the upper bound on
$m_{\rm LSP}$.

\begin{figure}
\centering
\includegraphics[height=4.1in,angle=-90]
{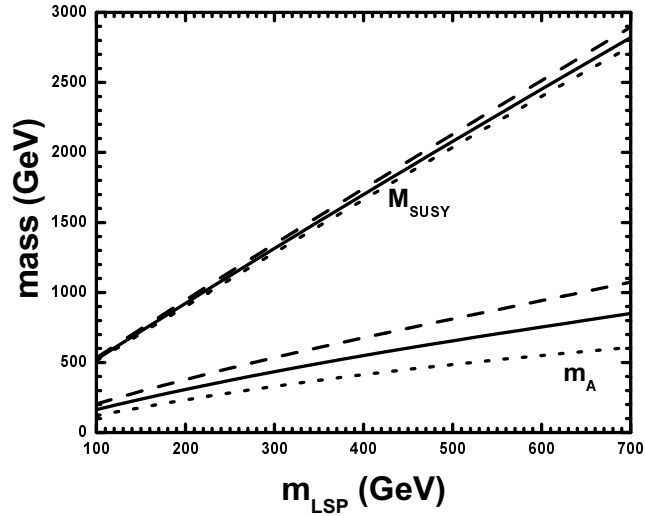}
\caption{The mass parameters $m_A$ and
$M_{\rm SUSY}$ as functions of $m_{\rm LSP}$
for $\mu>0$, $A_0=0$,
$\Delta_{\tilde\tau_2}=1$,
$\alpha_s(M_Z)=0.1185$, and with
$m_b(M_Z)=2.684~{\rm GeV}$ (dashed lines),
$3.092~{\rm GeV}$ (dotted lines), or
$2.888~{\rm GeV}$ (solid lines).}
\label{figc}
\end{figure}

\item For low $\Delta_{\tilde\tau_2}$'s,
bino-stau coannihilations \cite{ellis2} take
over leading to a very pronounced reduction of
the LSP relic abundance $\Omega_{\rm LSP}h^2$,
thereby enhancing the upper limit on
$m_{\rm LSP}$. So, we obtain the almost
horizontal tail of the allowed region in
Fig.~\ref{figa}.

\end{itemize}

\par
For $\mu>0$, $A_0=0$, $\alpha_s(M_Z)=0.1185$
and $m_b(M_Z)=2.888~{\rm GeV}$, we find the
following allowed ranges of parameters:
\begin{eqnarray}
&&
176~{\rm GeV}\stackrel{<}{_{\sim}} m_{\rm LSP}
\stackrel{<}{_{\sim}} 615~{\rm GeV},\quad
0\stackrel{<}{_{\sim}}\Delta_{\tilde\tau_2}
\stackrel{<}{_{\sim}} 1.8,
\nonumber \\
&& 58\stackrel{<}{_{\sim}}\tan\beta
\stackrel{<}{_{\sim}} 59,\quad 0.14
\stackrel{<}{_{\sim}} c\stackrel{<}{_{\sim}}
0.17.
\end{eqnarray}
The splitting between the bottom
(or tau) and top Yukawa coupling constants $\delta h
\equiv -(h_b-h_t)/h_t=(h_\tau-h_t)/h_t=2c/(1+c)$
ranges between 0.25 and 0.29.

\subsection{The New Shifted Hybrid Inflation}
\label{sec:shift}

\par
It is interesting to note that our SUSY GUT
model gives rise \cite{jean2} naturally to a
modified version of shifted hybrid inflation
\cite{jean}. Hybrid inflation \cite{linde},
which is certainly one of the most promising
inflationary scenarios, uses two real scalars:
one which provides the vacuum energy density
for driving inflation and a second which is
the slowly varying field during inflation. This
scheme, which is naturally incorporated
\cite{hybrid} in SUSY GUTs (for an updated
review, see Ref.~\cite{senoguz}), in its
standard realization has the following property
\cite{pana1}: if the GUT gauge symmetry
breaking predicts topological defects such as
magnetic monopoles \cite{monopole}, cosmic
strings \cite{strings}, or domain walls
\cite{wall}, these defects
are copiously produced at the end of inflation.
In the case of monopoles or walls, this leads
to a cosmological catastrophe \cite{kibble}.
The breaking of the $G_{\rm PS}$ symmetry
predicts the existence of doubly charged
monopoles \cite{laz3}. So, any PS SUSY GUT
model
incorporating the standard realization of SUSY
hybrid inflation would be unacceptable. One way
to remedy this is to invoke \cite{thermal1}
thermal inflation \cite{thermal2} to dilute the
primordial monopoles well after their
production. Alternatively, we can construct
variants of the standard SUSY hybrid
inflationary scenario such as smooth
\cite{pana1} or shifted \cite{jean} hybrid
inflation which do not suffer from the monopole
overproduction problem. In the latter scenario,
we generate \cite{jean} a shifted inflationary
trajectory so that $G_{\rm PS}$ is already
broken during inflation. This could be achieved
\cite{jean} in our SUSY GUT model even before the
introduction of the extra Higgs superfields,
but only by utilizing non-renormalizable terms.
The inclusion of $h^\prime$ and
$\bar{h}^\prime$ does not change this
situation. The inclusion of $\phi$ and
$\bar\phi$, however, very naturally gives rise
\cite{jean2} to a shifted path, but now with
renormalizable interactions alone.

\section{Conclusions}
\label{sec:concl}

\par
We showed that particle physics provides us
with a number of candidate particles out of
which the CDM of the universe can be made.
These particles are not invented solely for
explaining the CDM, but they are naturally
there in various particle physics models.
We discussed in some detail the major
candidates which are the axion, the lightest
neutralino, the axino, and the gravitino. The
last three particles exist only in SUSY
theories and can be stable provided that they
are the LSP.

\par
The axion is a pseudo Nambu-Goldstone boson
associated with the spontaneous breaking of a
PQ symmetry. This is a global anomalous
${\rm U}(1)$ symmetry invoked to solve the
strong $CP$ problem. It is, actually, the most
natural solution to this problem which is
available at present. The axions are extremely
light particles and are generated at the QCD
phase transition carrying zero momentum. We
argued that these particles can easily provide
the CDM in the universe. However, if the PQ
field emerges with non-zero value at the end of
inflation, they lead
to isocurvature perturbations, which, for
superheavy inflationary scales, are too
strong to be compatible with the recent results
of the WMAP satellite on the CMBR anisotropies.

\par
The most popular CDM candidate is, certainly,
the lightest neutralino which is present in all
SUSY models and can be the LSP for a wide range
of parameters. We considered it within the
simplest SUSY framework which is the MSSM whose
salient properties were summarized. We used
exclusively the constrained version of
MSSM which is known as CMSSM and is based on
universal boundary conditions. In this case,
the lightest neutralino is an almost pure bino,
whereas the NLSP is the lightest stau. We
sketched the calculation of the neutralino
relic abundance in the universe paying
particular attention not only to the neutralino
pair annihilations, but to the neutralino-stau
coannihilations too. It is very important for
the accuracy of the calculation to treat poles
and final-state thresholds properly and include
the one-loop QCD corrections to the Higgs boson
decay widths and the fermion masses. We find
that two effects help us reduce the
neutralino relic abundance and satisfy the WMAP
constraint on CDM: the resonantly enhanced
neutralino pair annihilation via an $A$-pole
exchange in the s-channel, which appears in the
large $\tan\beta$ regime, and the strong
neutralino-stau coannihilation, which is
achieved when these particles are almost
degenerate in mass.

\par
The axino, which is the SUSY partner of the
axion, can also be the LSP in many cases since
its mass is a strongly model-dependent
parameter in the CMSSM. It is produced
thermally by 2-body scattering or decay
processes in the thermal bath, or
non-thermally by the decay of sparticles which
are already frozen out of thermal equilibrium.
For small axino masses, TP is more important
yielding a very narrow favored region in the
parameter space. For larger axino masses,
however, NTP is more efficient and the favored
region in the parameter space becomes
considerably wider. One finds that, in the case
of the CMSSM, almost
any point on the $m_0-M_{1/2}$ plane can be
allowed by axino CDM considerations. The
required reheat temperatures though are quite
small ($\stackrel{<}{_{\sim}}{\rm few}\times
100~{\rm GeV}$).

\par
The mass of the gravitino is a practically
free parameter in the CMSSM. So, the gravitino
can easily be the LSP and, in principle,
contribute to the CDM of the universe. It is
produced thermally by 2-body scattering
processes in the thermal bath as well as
non-thermally by the decay of the NLSP, which
can be either the neutralino or the stau. In
contrast to the axino case, however, the NLSP
can now have quite a long lifetime. The
electromagnetic showers resulting from the NLSP
decay can destroy the successful predictions of
BBN. So, we obtain strong constraints which
allow only very limited regions of the
parameter space of the CMSSM. As it turns out,
NTP in these regions is not efficient enough to
account for CDM. We can, however, make these
regions cosmologically favored by raising
$T_{\rm r}$ to enhance TP of gravitinos.

\par
We studied the CMSSM with $\mu>0$ and $A_0=0$
applying a YQUC which originates from a PS SUSY
GUT model. This condition yields an adequate
deviation from YU which allows an acceptable
$m_b(M_Z)$. We, also,
imposed the constraints from the CDM in the
universe, $b\rightarrow s\gamma$,
$\delta\alpha_\mu$ and $m_h$. We found that
there exists a wide and natural range of CMSSM
parameters which is consistent with all the
above constraints. The parameter $\tan\beta$
ranges between about 58 and 59 and the
asymptotic splitting between the bottom (or
tau) and the top Yukawa coupling constants
varies in the range $25-29\%$ for central
values of $m_b(M_Z)$ and $\alpha_s(M_Z)$.
The predicted LSP mass can be as low as about
$176~{\rm GeV}$. Moreover, the model resolves
the $\mu$ problem of MSSM, predicts
stable proton, generates the baryon asymmetry
of the universe via primordial leptogenesis,
and gives rise to a new version of shifted
hybrid inflation which is based solely on
renormalizable interactions.

\section*{Acknowledgements}
We thank L. Roszkowski, P. Sikivie, and F.D.
Steffen for useful suggestions. This work was
supported by European Union under the contract
MRTN-CT-2004-503369 as well as the Greek
Ministry of Education and Religion and the
EPEAK program Pythagoras.



\def\ijmp#1#2#3{{Int. Jour. Mod. Phys.}
{\bf #1},~#3~(#2)}
\def\plb#1#2#3{{Phys. Lett. B }{\bf #1},~#3~(#2)}
\def\zpc#1#2#3{{Z. Phys. C }{\bf #1},~#3~(#2)}
\def\prl#1#2#3{{Phys. Rev. Lett.}
{\bf #1},~#3~(#2)}
\def\rmp#1#2#3{{Rev. Mod. Phys.}
{\bf #1},~#3~(#2)}
\def\prep#1#2#3{{Phys. Rep. }{\bf #1},~#3~(#2)}
\def\prd#1#2#3{{Phys. Rev. D }{\bf #1},~#3~(#2)}
\def\npb#1#2#3{{Nucl. Phys. }{\bf B#1},~#3~(#2)}
\def\npps#1#2#3{{Nucl. Phys. B (Proc. Sup.)}
{\bf #1},~#3~(#2)}
\def\mpl#1#2#3{{Mod. Phys. Lett.}
{\bf #1},~#3~(#2)}
\def\mpla#1#2#3{{Mod. Phys. Lett. A}
{\bf #1},~#3~(#2)}
\def\arnps#1#2#3{{Annu. Rev. Nucl. Part. Sci.}
{\bf #1},~#3~(#2)}
\def\sjnp#1#2#3{{Sov. J. Nucl. Phys.}
{\bf #1},~#3~(#2)}
\def\jetp#1#2#3{{JETP Lett. }{\bf #1},~#3~(#2)}
\def\app#1#2#3{{Acta Phys. Polon.}
{\bf #1},~#3~(#2)}
\def\rnc#1#2#3{{Riv. Nuovo Cim.}
{\bf #1},~#3~(#2)}
\def\ap#1#2#3{{Ann. Phys. }{\bf #1},~#3~(#2)}
\def\ptp#1#2#3{{Prog. Theor. Phys.}
{\bf #1},~#3~(#2)}
\def\apjl#1#2#3{{Astrophys. J. Lett.}
{\bf #1},~#3~(#2)}
\def\n#1#2#3{{Nature }{\bf #1},~#3~(#2)}
\def\apj#1#2#3{{Astrophys. J.}
{\bf #1},~#3~(#2)}
\def\anj#1#2#3{{Astron. J. }{\bf #1},~#3~(#2)}
\def\mnras#1#2#3{{MNRAS }{\bf #1},~#3~(#2)}
\def\grg#1#2#3{{Gen. Rel. Grav.}
{\bf #1},~#3~(#2)}
\def\s#1#2#3{{Science }{\bf #1},~#3~(#2)}
\def\baas#1#2#3{{Bull. Am. Astron. Soc.}
{\bf #1},~#3~(#2)}
\def\ibid#1#2#3{{\it ibid. }{\bf #1},~#3~(#2)}
\def\cpc#1#2#3{{Comput. Phys. Commun.}
{\bf #1},~#3~(#2)}
\def\astp#1#2#3{{Astropart. Phys.}
{\bf #1},~#3~(#2)}
\def\epjc#1#2#3{{Eur. Phys. J. C}
{\bf #1},~#3~(#2)}
\def\nima#1#2#3{{Nucl. Instrum. Meth. A}
{\bf #1},~#3~(#2)}
\def\jhep#1#2#3{{J. High Energy Phys.}
{\bf #1},~#3~(#2)}
\def\jcap#1#2#3{{J. Cosmol. Astropart. Phys.}
{\bf #1},~#3~(#2)}
\def\apjs#1#2#3{{Astrophys. J. Suppl.}
{\bf #1},~#3~(#2)}
\def\lnp#1#2#3{{Lect. Notes Phys.}
{\bf #1},~#3~(#2)}
\def\jpa#1#2#3{{J. Phys. A}
{\bf #1},~#3~(#2)}
\def\ppnp#1#2#3{{Prog. Part. Nucl. Phys.}
{\bf #1},~#3~(#2)}
\def\jetpsp#1#2#3{{JETP (Sov. Phys.)}
{\textbf {#1}},~#2~(#3)}

\end{document}